\documentclass[apjl]{emulateapj}
\usepackage{mathptmx}

\begin{document}

\title{A Kennicutt-Schmidt Law for Intervening Absorption Line Systems}
\author{Doron Chelouche\altaffilmark{1} and David V. Bowen\altaffilmark{2}}
\altaffiltext{1} {Department of Physics, Faculty of Natural Sciences, University of Haifa, Haifa, 31905, Israel; doron@sci.haifa.ac.il}
\altaffiltext{2} {Department of Astrophysical Sciences, Peyton Hall, Princeton University, 
Princeton, NJ 08544, USA; dvb@astro.princeton.edu}
\shortauthors{Chelouche \& Bowen}
\shorttitle{A Kennicutt-Schmidt Law for Intervening Systems}

\begin{abstract}

We argue that most strong intervening metal absorption line systems, where the rest equivalent width of the \ion{Mg}{2}\,$\lambda 2796$ line is $>0.5$\AA, are interstellar material in, and outflowing from, star-forming disks. We show that a version of the Kennicutt-Schmidt law is readily obtained if the \ion{Mg}{2}\ equivalent widths are interpreted as kinematic broadening from absorbing gas in outflowing winds originating from star-forming galaxies.  Taking a phenomenological approach and using a set of observational constraints available for star-forming galaxies, we are able  to account for the density distribution of strong \ion{Mg}{2}\ absorbers over cosmic time. The association of intervening material with star-forming disks naturally explains the metallicity and dust content of strong \ion{Mg}{2}\ systems as well as their high \ion{H}{1}\ column densities, and does not require the advection of metals from compact star-forming regions into the galaxy halos to account for the observations. We find that galaxies with a broad range of luminosities can give rise to absorption of a given rest-equivalent width, and discuss possible observational strategies to better quantify true galaxy-absorber associations and further test our model.   We show that the redshift evolution in the density of absorbers closely tracks the star formation history of the universe and that strong intervening systems can be used to directly probe the physics of both bright and faint galaxies over a broad redshift range. In particular, in its simplest form, our model suggests that many of the statistical properties of star-forming galaxies and their associated outflows have not evolved significantly since $z\sim 2$. By identifying strong intervening systems with galaxy disks and quantifying a version of the Kennicutt-Schmidt law that applies to them, a new probe of the interstellar medium is found which provides complementary information to that obtained through emission studies of galaxies. Implications of our results for galaxy feedback and enrichment of the intergalactic medium are discussed.

\end{abstract}

\keywords{
ISM: jets and outflows --- 
galaxies: evolution --- formation --- intergalactic medium ---
quasars: absorption lines ---
stars: formation
}

\section{Introduction}

Quasar spectra often reveal the presence of intervening metal absorption systems that span the entire observable redshift range between the observer and the source \citep{bu66}. A possible association of such absorption systems with intervening galaxy disks was first suggested by \citet{wa67}.  An alternative explanation associating such systems with large (compared to the size of visible disks) halos of galaxies was considered by \citet{bs69}, who paved the way to the use of intervening absorption systems as probes of the large scale environment of galaxies \citep{wey78}.  Despite enormous efforts over the last three decades to determine whether absorption arises in galaxy disks or halos, the question remains largely unresolved. This is particularly true for the \ion{Mg}{2}\,$\lambda \lambda 2796, 2803$ absorption line systems; these absorbers can be detected at relatively low redshifts ($z \ga 0.2$) in the optical spectra of quasars, so understanding their origin has been a more tractable problem for modern instrumentation, as we now discuss. 

Initial follow-up imaging and spectroscopy of galaxies in the fields of quasars known to show \ion{Mg}{2}\ absorption confirmed the association of strong (with rest-equivalent width $W_0\gtrsim 1$\AA) intervening systems with galaxies whose luminosities are in the range $10^{-2}L^\star-10\,L^\star$ \citep{be86,ste95}. Nevertheless, a clear picture connecting absorbers and galaxies has yet to emerge \citep{yo82,be91,bec92,ste94,ste97,ste02,tr05,kac07,ch08}. Much progress has been made in characterizing the properties of absorption line systems from studying the spectra of large numbers of QSOs discovered in wide-area sky surveys \citep{nes05,pr06b,yo06}. Following on from the earliest analysis of the redshift distribution of absorption lines \citep{lan87,tyt87,sar88} it is now well documented that the cosmic density of absorbers per unit redshift and unit rest-equivalent width, as a function of $W_0$,  $\partial^2 N/\partial z \partial W_0$, follows an exponential distribution \citep{nes05,pr06b}, but how this distribution is related to the association of absorption lines with galaxies, is unknown.  Recent statistical analyses reveal that  strong \ion{Mg}{2}\  systems are dusty \citep{yo06,me08} and that the dust-to-gas ratio of moderate redshift systems is consistent with local interstellar material (ISM) values \citep[but may be substantially lower at higher redshifts \citep{pet96}]{me09}. This suggests that the gas composition of such systems is not considerably different from solar values \citep{yo06,me09}. There are recent indications that strong systems are associated with star formation activity \citep{bou06,bou07l,wi07,zi07,me10} yet the nature of this connection is not well understood.

Despite the recent  progress in the phenomenology of intervening absorption systems, little is known with confidence about the physical state and origin of the gas \citep[and references therein]{ch05,c08}. Different models attribute the absorption to either galactic winds [e.g., \citet{bou08,op08}], or to cool clumps condensing out from the virialized gaseous halos of galaxies [e.g., \cite{mo96,ma04,ka08}] related, perhaps, to infalling material \citep{ti08,ker09}. Alternative explanations relate some systems to gravitationally bound minihalos \citep{kep99,ster02,gn04} or  to galaxy disks \citep{wa67,bo95,pr97,ca96,pr98,zw05,zw08}. While some models are qualitatively consistent with some observational constraints [e.g., \citet{sre94,c08,ti08}], many of the recently discovered statistical properties of \ion{Mg}{2}\ systems have not been addressed by theoretical works, and the following questions remain: What is the origin of strong metal systems? What sets their observational signatures? How do they relate to galactic and intergalactic medium (IGM) processes? In this we attempt to shed light on some of those questions by establishing a firmer connection between strong intervening systems and star forming galactic disks. 

The paper is organized as follows: section 2 qualitatively argues for a Kennicutt-Schmidt like star formation law \citep{sh59,ke89,ke98} for strong intervening \ion{Mg}{2}\ systems.  A more quantitative approach is taken in \S3 where a phenomenological model is formulated, which links the properties of star-forming galaxies to those of absorbers. The results of our model fits to the observations are outlined in \S4. We discuss the implications of our results for the study of strong intervening systems and galaxy disks across cosmic time in \S5. A summary follows in \S6.

\section{A Kennicutt-Schmidt law for strong \ion{Mg}{2} Absorbers}

Perhaps the best known property of star forming galaxy disks is that they obey the Kennicutt-Schmidt law for star formation \citep{sh59,ke89,ke98} whereby the star formation rate per unit area scales with the mean column density of the disk to some power\footnote{We note, however, possible deviations from a simple powerlaw at low column densities and star formation rates \citep{bi08}; see also \S5.8}. The underlying physical mechanism behind this phenomenological scaling is not fully understood and considerable theoretical and observational effort has been invested in identifying its origin \citep{si97,kra03,kr05,bi08,le08,kr09,gen10,gk10}. 

Recent spectroscopic observations reveal that many star forming galaxies give rise to outflows of cool gas from their centers \citep{hec93,hec01,ma05,rup05,vei05,tr07,we08,ru09}. Furthermore,  a relation exists between the outflow velocity and the star formation rate \citep{ma05,we08,ru09}. The physical nature of such outflows and their prevalence in the galaxy population as a whole, are currently poorly constrained. Nevertheless, it has been suggested that such outflows can provide the necessary negative feedback effects that determine and maintain the observed star formation rates observed in galaxies \citep{st99,sha03,mu05}, and as a means for enriching the IGM \citep[and references therein]{dal07,op08}.

\begin{figure}
\plotone{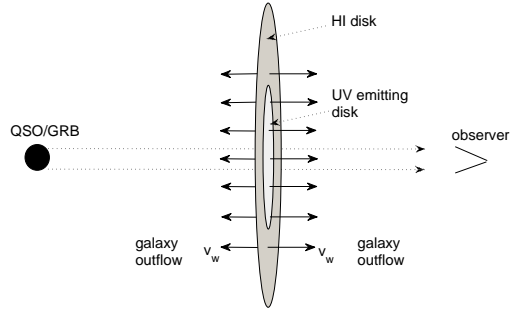}
\caption{A schematic view of a wind launched from a galactic disk (here we neglect inclination effects). The length of the arrows does {\it not} reflect the extent of the outflow above the disk and is for illustrative purposes only. The cross-section of the wind may be smaller than that of the \ion{H}{1}\ disk but is unlikely to be much smaller than that of the UV emitting disk (see text). The spectrum of background luminous objects (whose light is not significantly attenuated by extinction), such as quasars and $\gamma$-ray bursts (GRBs) will show absorption systems whose velocity dispersion is determined by the wind kinematics on the approaching and the receding side of the disk. UV observations of galaxies in which star forming regions are used as the background UV source probe only the approaching side of the disk outflow. We neglect contributions to the absorption signal from infalling material. Inclination effects (not shown) may be important and are incorporated into our model using data from \citet{ma05}.}
\label{pic}
\end{figure}

As mentioned above, there is mounting evidence that strong \ion{Mg}{2}\ absorbers ($W_0>1$\AA) are connected to star formation activity and, as such, could be related to the gaseous outflows associated with it \citep{nor96,pr06b,bou07l}. Here we argue that strong \ion{Mg}{2}\ absorbers {\it are}, in fact, star forming galactic disks and their associated outflows. To this end we note two observationally robust scalings: the relation between the (geometric) mean \ion{H}{1}\ column density of strong \ion{Mg}{2}\ absorbers and $W_0$ such that \citep{me09},
\begin{equation}
N_{\rm HI} = (3\pm0.5)\times 10^{19} W_0^{1.7\pm0.3}\,{\rm cm^{-2}},
\label{nhi}
\end{equation}
and the relation between the (maximum) velocity of galactic winds, $v_w$, and the star formation rate (SFR, denoted by $\Upsilon$) such that 
\begin{equation}
v_w \simeq 70\xi_w\, \left ( \frac{\Upsilon}{{\rm 1 M_\odot~yr^{-1}}} \right )^{0.35\pm0.05}\,{\rm km~s^{-1}}.
\label{vw}
\end{equation}
Here, $\xi_w=1$ for \ion{Na}{1}\ lines \citep{ma05} and $\xi_w\lesssim 3$ for \ion{Mg}{2}\  lines, as statistically estimated from the stacked spectra of a sample of star-forming DEEP2 galaxies \citep{we08}. To clarify we note that $\xi_w>1$ does not necessarily mean that \ion{Mg}{2}\ ions travel faster than \ion{Na}{1}\ ions and that it may be that observational effects, such as optical depth or partial filling by emission are important \citep{ma09}.

Here we assume, with little justification at this stage, that outflow kinematics are responsible for the line broadening of strong \ion{Mg}{2} systems. In particular, we consider a case in which the rest equivalent width of strong \ion{Mg}{2}\ troughs and the velocities of outflows are related by 
\begin{equation}
W_0\simeq2\left( \frac{v_w}{100\,{\rm km\,s^{-1}}} \right ){\rm \AA}.
\label{w0}
\end{equation}
This proportionality between the kinematic broadening and $W_0$ is justified for strong \ion{Mg}{2}\ systems whose absorption troughs are saturated, and $W_0$ mirrors the velocity dispersion of the gas. Since $v_w$ is a measure of the outflow velocity towards the center of a galaxy, we add the proportionality factor of 2 to account for a background object sightline intercepting both the receding and advancing sides of a symmetric outflow (see Fig. \ref{pic}). This model neglects the contribution of infalling material [e.g., cool filaments or galactic fountains \citep{br80,bro09}] whose covering factor is expected to be small \citep{sat09} and the velocity spread moderate. It also neglects contribution to the velocity field from galactic rotation which is expected to be small for thin disks, as determined from current data \citep{zw08}. Inclination effects will be treated in \S3.

Combining the above, seemingly unrelated, relations for $N_{\rm HI}(W_0)$ and $v_w(\Upsilon)$ using equation \ref{w0}, we obtain an expression relating the SFR to the \ion{H}{1}\ column density, with the following form:
\begin{equation}
\Upsilon \simeq  (0.1\pm0.03)\times \left ( \frac{N_{\rm HI}}{10^{20}\,{\rm cm^{-2}}} \right )^{1.7\pm0.3}\left ( \frac{\xi_w}{3} \right )^{-2.9\pm0.4}\,{\rm M_\odot~yr}^{-1}.
\label{sk0}
\end{equation}
This relationship between the SFR of a \ion{Mg}{2}\ host galaxy and the \ion{H}{1}\ column density of the absorbing gas, qualitatively has the same functional form as the Kennicutt-Schmidt law for star formation in galaxies \citep{ke98,bou07,gen10}. An analogous scaling is obtained if, instead of using equation \ref{nhi}, one uses the relation between the dust column density and $W_0$ found by \cite{me08}\footnote{\citet{me08} find that the reddening $E(B-V)$ due to intervening systems' dust and $W_0$ are related such that $E(B-V) \propto W_0^{1.9\pm0.2}$. Assuming the reddening is a proxy for the dust column and a constant dust-to-gas ratio, a completely analogous relation to equation \ref{nhi} is obtained.}.

Let us now crudely estimate a Kennicutt-Schmidt relation from equation \ref{sk0}: we recall that $W_0\gtrsim 1$\AA\ systems are thought to be associated with $L^\star$ galaxies \citep{ste95}, whose disk radius we take to be $R_{\rm eff} \sim 10$\,kpc [qualitatively consistent with the relation reported by \citet{cou07}]. An order-of-magnitude estimate for the star formation surface density is therefore $\Sigma_\star \equiv \Upsilon/(\pi R_{\rm eff}^2)\sim 3\times 10^{-4} N_{\rm HI,20}^{1.7\pm0.3} (\xi_w/3)^{-2.9\pm0.4}\,{\rm M_\odot~yr^{-1}~kpc^{-2}}$ (where $N_{{\rm HI},20} \equiv N_{\rm HI}/10^{20}\,{\rm cm^{-2}}$). We hence obtain a surprisingly good match with the galactic Kennicutt-Schmidt law \citep{ke98}. It is important to note that the uncertainties on the proportionality factor in this relation are large: for example, if the typical galactic disk radius giving rise to strong systems is only 3\,kpc large (as would be the case if the hosts of most strong systems are fainter; see below) then the Kennicutt-Schmidt law obeyed by absorbers implies a higher star formation rate for the same \ion{H}{1}\ column density compared to galaxies. We return to these important issues when considering a more quantitative version of this relation in \S5.8.

Obtaining the form of the Kennicutt-Schmidt law by combining pure absorber and galaxy phenomenology raises the interesting possibility that (a) sight-lines through most strong \ion{Mg}{2}\ absorbers probe star forming disks [as opposed to e.g., condensed halo gas; \citep{mo96}] and that (b) the broadening of \ion{Mg}{2}\ troughs is largely due to disk wind kinematics rather than, for example, the velocity dispersion of absorbing clouds in galactic halos or the rotation speed of gas in galactic disks. This intimate absorber-disk association also naturally explains the large \ion{H}{1} column densities associated with strong intervening systems, which are comparable to those observed through disks. It also explains the high dust-to-gas ratios in those systems \citep{me09}, as well as their association with star forming regions. 

\section{The Model}

If most strong \ion{Mg}{2}\ systems probe galaxy disks then their statistical properties should be accounted for by those of galaxies. In particular, the exponential density distribution of such absorption systems as a function of $W_0$, $\partial^2 N/\partial z\partial W_0$  should be describable by the following expression\footnote{We assume a flat universe with $\Omega_\Lambda=0.7$, $\Omega_M=0.3$, and $H_0=70\,{\rm km~s^{-1}~Mpc^{-1}}$.}:
\begin{equation}
\frac{\partial^2N(W_0,z)}{\partial z \partial W_0} =\frac{c}{H_0}\frac{(1+z)^2}{E(z)} \int_{L_{\rm min}}^{L_{\rm max}} dL\phi_z(L) \sigma_z(L) P_z(W_0 \vert L)
\label{conv}
\end{equation}
where $E(z)=\sqrt{\Omega_M (1+z)^3+\Omega_\Lambda}$, $\phi_z(L)$ is the redshift-dependent luminosity function, $L_{\rm min}~(L_{\rm max})$ is the minimum (maximum) luminosity of galaxies that give rise to strong \ion{Mg}{2}\ systems, and $\sigma_z(L)$ is the luminosity and redshift-dependent geometrical cross-section for detecting a \ion{Mg}{2}\ absorption around a galaxy of luminosity $L$. $P(W_0\vert L)$ if the probability of detecting a system with a rest equivalent width $W_0$ around a  galaxy of  luminosity $L$\footnote{Unless otherwise specified, the following normalization holds: $\int P(x\vert y)dx=1$}. This probability  may be written as  
\begin{equation}
P_z(W_0 \vert L)= \int d\Upsilon P_{z}(W_0 \vert \Upsilon) P_{z}(\Upsilon \vert L)  ,
\end{equation} 
where $P_z(\Upsilon \vert L)$ is the probability of having a certain SFR, $\Upsilon$, for galaxy of luminosity $L$, and $P_z(W_0 \vert \Upsilon)$ is the probability of having a certain $W_0$ - i.e., outflow velocity - for a galaxy with a star formation rate $\Upsilon$. For simplicity we assume that there are no hidden correlations among the various variables apart from those explicitly stated here. Whether this holds in realistic systems is unclear, but assessing the degree to which any hidden correlations may change our results is beyond the scope of this paper. 

\begin{figure}
\plotone{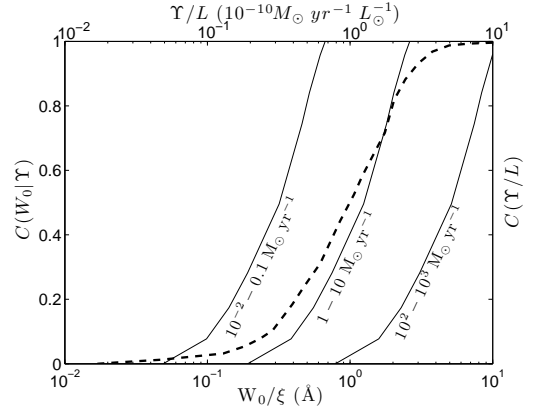}
\caption{The observationally derived cumulative probabilities [$C(x) \equiv \int_0^{x} P(x') dx'$] used in this paper. $C(W_0 \vert \Upsilon)$ (solid lines for several SFR values) was obtained from \cite{ma05} (after conversion to $W_0$; see text). Clearly, an order of magnitude range in $W_0$ may be obtained for any given SFR bin [see also \citet{ma05} where this holds also for narrow bin sizes]. Generally, however, smaller SFR result in lower outflow velocities, hence $W_0$ values. $C(\Upsilon/L)$ (dashed line) shows the specific SFR distribution as determined from \cite{ja08}. Clearly, a broad SFR range can be obtained for any given galaxy.}
\label{Ps}
\end{figure}

The model defined above seeks to connect the observable properties of absorbers to those of their host galaxies. Unfortunately, the properties of galaxies and absorption line systems are well measured only over particular redshift intervals, and these intervals do not always overlap. The properties of galaxies are best understood at very low redshift, but the characteristics of UV absorption line systems have been determined largely at intermediate and high redshift.  For example, as we  shall see below, faint galaxies have a potentially important contribution to the cosmic density of strong systems and their properties are known only locally. We can therefore proceed in one of two ways: we can (a) extrapolate  the local properties of (faint) galaxies to higher redshifts, or (b) extrapolate the properties of absorbers to low redshifts. The first approach requires the extrapolation of the locally determined galaxy luminosity function, $\sigma_{z=0} (L),~P_{z=0}(\Upsilon \vert L),~{\rm and}~ P_{z=0}(W_0 \vert \Upsilon)$ to higher $z$. In contrast, the second approach requires only that we extrapolate ${\partial ^2 N(W_0)}/{\partial z \partial W_0}$ to low redshifts. Furthermore, with the installation of the {\it Cosmic Origin Spectrograph} (COS) on the {\it Hubble Space Telescope} (HST), the properties of low redshift absorption lines are likely to be determined (at least for $W_0\gtrsim 1$\AA\ systems) in the near future. Conversely, many of the properties of faint galaxies at high redshifts are likely to remain uncertain for some time. 

With the above in mind, we restrict our analysis to $z\sim 0$ and will henceforth drop the $z-$dependence from our derived values and distributions (we return to the effects of redshift evolution in \S5.5).  Specifically, we take a Schechter form for the luminosity function of galaxies that give rise to metal absorption systems
\begin{equation}
\phi(L)dL=f_L \phi^\star \left ( \frac{L}{L^\star} \right )^\alpha e^{-L/L^\star}\frac{dL}{L^\star}
\end{equation}
We adopt the parameterization of \citet{ja08} who carried out a census of star-forming galaxies in the local universe, and found: $\alpha\simeq -1.4 $, $\phi^\star \sim 3\times 10^{-3}\,{\rm Mpc}^{-3}$ ($L^\star\sim 1.4\times 10^{10}L_\odot$, with $L_\odot$ being the solar luminosity). We note that this particular parameterization for the luminosity function of galaxies is one out of many, and somewhat different, parameter values are reported in the literature depending on the selection criteria and filter used \citep{maz98,zw01,no02,bl03}. As we do not know whether all star-forming disk galaxies give rise to \ion{Mg}{2}\ systems,  we define $f_L$  to be the fraction of galaxies that cause \ion{Mg}{2}\ absorption and allow it to be smaller than unity. We neglect a possible dependence of $f_L$ on the galaxy luminosity. We take $L_{\rm max} \to \infty$ implying that the model considers all galaxies with luminosity, $L>L_{\rm min}$ as contributing to the population of \ion{Mg}{2}\ absorbers. It turns out that, due to the exponentially small number of very luminous galaxies, the particular $L_{\rm max}$ chosen has little effect on the predicted number density of $W_0 \gtrsim 1$\AA\ absorbers. That said, the particular choice of $L_{\rm max}$ may affect model predictions concerning the number density of the strongest systems, which are rare and their statistics rather poorly determined.

The cross-section for \ion{Mg}{2}\ absorption from galaxy disks, $\sigma(L)=\pi R_{\rm MgII}(L)^2$ [with $R_{\rm MgII}(L)$ being the effective radius of the absorbing region], is unknown, as is its dependence on galaxy luminosity. We assume that this cross-section scales with that of the \ion{H}{1}\ disk, $\sigma_{\rm HI}$, as deduced from 21\,cm emission down to a column density limit of $\gtrsim 10^{19}\,{\rm cm^{-2}}$ \citep{zw05}.  We therefore define 
\begin{equation}
\sigma(L)=\xi_\sigma \sigma_{\rm HI} (L^\star) \left ( \frac{L}{L^\star} \right)^\gamma
\label{sigma}
\end{equation}
with $\gamma$ being of order unity, which is qualitatively consistent with the observed local scaling of \ion{H}{1}\ disk cross-sections [see Fig. 10 of \citet{zw05}, which shows a constant probability for detecting absorbers through disks spanning a range of luminosities]. For infrared galaxy disks $\gamma\sim 0.7$ \citep{cou07} while in the optical $\gamma \sim 0.6$ \citep{dej00}. Interestingly,  a similar scaling has been recently reported between the effective radius of \ion{Mg}{2}\ absorbing envelopes of galaxies and their luminosity \citep{ch10}.  In equation \ref{sigma}, $\xi_\sigma$ stands for the relative size of the \ion{Mg}{2}\ strong absorption cross-section compared to the \ion{H}{1}\ disk (assumed to be luminosity independent). This factor is currently poorly constrained and will be determined below from a fit of our model predictions to available absorption line data.

We next define the probabilities $P(\Upsilon \vert L)$ and $P(W_0 \vert \Upsilon)$ based on data from the literature. \citet[see their Fig. 3]{ja08} found no clear trend between the specific SFR (i.e., SFR per unit luminosity), as determined from H$\alpha$ luminosities, and galaxy luminosity; we determine $P(\Upsilon/L)$ from their data, and present its cumulative version in Figure \ref{Ps}.  The probability of having a certain rest equivalent width for the \ion{Mg}{2}\ line per given star formation rate, $P(W_0\vert \Upsilon)$, was determined from \citet{ma05} who measured the outflow velocity in  \ion{Na}{1}\ absorption lines as a function of the SFR. In doing so we assume that outflow kinematics, as observed along a sightline to a background object, traces global outflow properties \citep{ma05}, and neglect possible complications related to the effects of partial covering of the wind over the disk area \citet{ma09}. The translation from the outflow velocity of \ion{Na}{1}\ lines to $W_0$, using equation \ref{w0}, requires knowledge of $\xi_w$ (see \S2) whose value $\in [1,10]$ \citep[see also \S2]{ma05,tr07,we08} but is left here as a free parameter and will be determined from a fit of equation \ref{conv} to the available absorption line data.  We have used data pertaining to the high SFR bins  of \citet{ma05}, resulting in the best characterization of $P(W_0=\xi_w v_w \vert \Upsilon)$ for high SFRs (see her Fig. 6 for $\Upsilon > 100\,{\rm M_\odot~yr^{-1}}$).  We then assume that the deduced $P(W_0\vert \Upsilon)$ also holds for lower SFR up to a scaling factor such that the maximum velocity, hence $W_0$, in each star formation bin is $\propto \Upsilon^{\beta}$ with $\beta\simeq 0.3$ \citep[see our Fig. \ref{Ps}]{ma05,we08}. This is consistent with the \citet{ma05} results and is, presently, the only viable assumption given the quality of the data.  To summarize, our model thus far relies on observable properties of star-forming galaxies with only two quantities, $\xi_w$ and $\xi_\sigma$, both poorly constrained by observations and left as free parameters whose values will be determined by a fit to absorption line data.  

It is, currently, not well determined how inclination affects the observable properties of a galaxy outflow and any attempt to {\it directly} account for it in our model is subject to large uncertainties. While it is unlikely that an outflow is spherically symmetric, the details are obscure both observationally and theoretically [for a very recent work providing new insights into inclination issues see \citet{cheny10}]. That said, our model is unlikely to be considerably affected by such uncertainties since we use the {\it observationally} determined statistical properties of outflows from spectroscopic surveys of star forming galaxies. Such studies probe galaxies with a broad range of inclinations to our sight line, hence the effects of inclination already appears in the data and enter our model, by definition, via $P(W_0 \vert \Upsilon)$. 

We predict the number density of absorption line systems with a given equivalent width, ${\partial ^2 N(W_0,z=0)}/{\partial z \partial W_0}$ from equation \ref{conv} using a Monte Carlo scheme: we first pick a galaxy whose luminosity, $L$, follows from the luminosity function. Given $L$, we use $P(\Upsilon \vert L)$ to statistically determine its star formation rate. The following step statistically determines $W_0$ from $P(W_0\vert \Upsilon)$. At this point, we have fully specified the statistical relation between galaxy and absorbing properties. This scheme is iterated to create a mock catalog of galaxies and their absorbing characteristics from which the statistical properties of the predicted absorbers may be deduced and compared to observations. The size of the sample is determined by the requirement that good statistics are realized for the strongest \ion{Mg}{2}\ absorbers, which are inherently rare.

\begin{figure}
\plotone{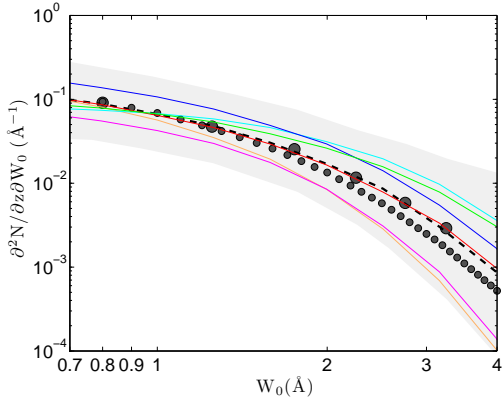}
\caption{The $\partial^2 N(z=0)/\partial z \partial W_0$ distribution with the normalization taken from \citet{pr06} and the uncertainty estimated to be a factor of $\sim 3$ for $W_0 \sim 1$, and somewhat larger for larger $W_0$ values (see \S4). The shape of the $\partial^2 N(z=0)/\partial z \partial W_0$ curve was determined from  \citet{nes05} by extrapolating their high redshift data (see their Fig. 10) to lower redshifts using their non-evolutionary curve models. Good agreement between the model and the data is shown in black dashed line and its parameters are: $\delta=\alpha+\gamma=-1.4+0.7=-0.7, ~\beta=0.3,~\xi_w=1.6,~R_{\rm MgII}(L^\star)=40\,{\rm kpc},\phi^\star=3\times 10^{-3}\,{\rm Mpc^{-3}},~L_{\rm min}=10^{-2}L^\star$ (model A; see table 1). Also shown are a few additional models, all of which are consistent with the data given the uncertainties, in colored curves (unless otherwise specified parameter values are identical to model A): $\xi_w=2$ (green), $\xi_w=1.1$ (orange), $\xi_w=1.1,~\delta=-0.4$ (magneta), $\beta=0.25$ (cyan), $R_{\rm MgII}(L^\star)=50$\,kpc (blue), and $\eta=0.1,~\xi_w=3.6$ (red; see \S4.1). Clearly, the origin of all strong systems with $W_0>0.7$\AA\ at $z\simeq 0$ is consistent  with local star forming disks. Diamonds mark our evolutionary model predictions for the signal at $z=0$ (see \S 5.5).}
\label{dndw}
\end{figure}

\subsection{The Kinematic Properties of Outflows Across Disks}

The properties of \ion{Mg}{2}\ winds across galaxy disks are poorly determined and only a characteristic velocity, $v_w$, over the entire UV emitting disk surface is measured. Whether or not outflows are thermally expanding or radiatively driven, it is likely that their kinematics will depend on their proximity to luminous or energetic sources in the disk such as star forming regions, dense stellar environments, or active galactic nuclei \citep[and references therein]{ma06,fuj09}. These are not distributed uniformly over the galaxy but rather cluster in the inner annuli of the disk \citep[and references therein]{le08,rm10}.  To allow for the (likely) possibility that gas kinematics varies across the disk surface of galaxies, we consider a general velocity profile of the form:
\begin{equation}
v_w(\rho)=\xi_w(\rho)v_w= \left (1+\frac{\rho}{\rho_0} \right )^\epsilon \xi_w v_w
\label{vwb}
\end{equation}
with $\rho$ being the impact parameter to the host's center and $\epsilon<0$ is a powerlaw index (with $\xi_w$ having the usual definition). We define the (projected) core radius within which the velocity profile flattens as some fraction of the radius of the \ion{Mg}{2}\ absorbing disk such that, $\rho_0(L)\equiv \eta R_{\rm MgII}(L)$  with $\eta$ assumed constant.  We choose $\epsilon=-0.5$ so that on scales $\rho \gg \rho_0$, an inverse square-root law is recovered mimicking, for example, the radial dependence of the escape velocity from a point mass \citep{cn05} or the terminal velocity of a radiation pressure driven cloud from a point source of a given luminosity \citep{cn01}. On scales $\rho < \rho_0$ galactic structure becomes important and the approximation of a point source breaks down leading to a flattening of the velocity profile.  Having very little information about the dynamics of galactic outflows, this is the simplest form for the velocity profile that is consistent with the data (see  \S4.1 for further details)\footnote{Not {\it all} radial forms are allowed: cuspy velocity profiles for which the velocity of the wind diverges as $\rho \to 0$ [as in the case, $v_w(\rho) \propto \rho^{\epsilon}$ for which $\eta \lll 1$] would be in disagreement with the data since, in this case, high $W_0$ values in the $\partial^2N/\partial z \partial W_0$ distribution would be determined by low luminosity objects rather than high luminosity ones. In this case,  model predictions will take on a powerlaw form at large $W_0$ values reflecting the geometrical cross-section for fast outflows rather than the  exponential decay of the luminosity function (see below).}  The inclusion of this part of the model in our Monte Carlo calculations is straightforward and requires us to randomly pick the impact parameter $\rho$ in addition to the host luminosity in order to predict $W_0$ for a specific sightline.   We present results for this extension to the model in \S4.1.

At this point, our model now relies on three free parameters: $\xi_\sigma(L^\star)$ [or $R_{\rm MgII}(L^\star)$], $\xi_w$, and $\eta$ all of which are poorly determined by observations.  In what follows we shall restrict our analysis to models in which $\eta$ is either $\gg 1$ (a constant wind velocity over the disk surface) or $\eta=0.1$ which roughly corresponds to the star-forming region of galaxies being enclosed within their inner few kpc.

\section{Results}

When comparing our model predictions to the data it is important to note that many of the parameters in equation \ref{conv} (e.g., $n_z,~\sigma_z$) are constrained by observations of galaxies at low redshifts and that their high-$z$ analogs may, in principle, have different values. Instead of attempting to predict how galaxy properties may have changed with redshift, we instead extrapolate $\partial^2 N/\partial z \partial W_0$ to redshifts of zero. As the statistics of strong \ion{Mg}{2}\ systems are only loosely determined at present times, we have used higher redshift data to estimate $\partial^2 N/\partial z \partial W_0$ at low-$z$. In particular, we used the non-evolutionary curves in Figure 10 of \citet{nes05} to estimate the relative normalization, i.e., the shape of $\partial^2 N/\partial z \partial W_0$, at different rest equivalent width slices used by the authors over the range $0.6<W_0<3.5$\AA.  The overall normalization of $\partial^2 N/\partial z \partial W_0$ at $z\simeq 0$ was chosen such that $\int_{ 1{\rm \AA} }^\infty dW_0 \partial^2 N(W_0,z=0)/\partial z \partial W_0=\partial N (W_0>1{\rm \AA}, z=0 ) / \partial z=0.05^{+0.1}_{-0.03}$, which is consistent with the extrapolations to $z=0$  based on the best-fit models of \citet[see their Table 4]{pr06b} to their high redshift data set. The quoted uncertainty qualitatively brackets the range in values predicted by the different extrapolation methods used in \citet{pr06b} and the measured value of $0.16^{+0.09}_{-0.05}$ reported by \citet[see his Fig. 7]{ch01} for $z\simeq 0.05$ systems. The number statistics of stronger systems ($W_0 > 2$\AA) is less secure at low redshifts, and our uncertainty for $W_0>2$\AA\ bins was chosen to include the factor between the predictions of the non-evolutionary models and the evolutionary curves of \citet[compare their Figs. 10 and 13]{nes05} for $z=0$. We neglect sample incompleteness effects arising from dust extinction at the high $W_0$ end of the distribution but note that these can result in an under-estimation of $\partial N(W_0> 3{\rm \AA},z\simeq 1)/\partial z$ by up to $\sim 20\%$ \citep{me08}. 

The numerical evaluation of equation \ref{conv}, for the case of a constant wind velocity over the disk surface, and its comparison to the observed number density of \ion{Mg}{2}\ lines, as extrapolated to $z=0$, is shown in Figure \ref{dndw}. (The solution for a model in which the wind varies as a function of $\rho$, with $\eta<1$, is discussed in \S4.1.) In this case, the fitting scheme involves a search in a two parameter space that is spanned by $\xi_w$ and $\xi_\sigma$ that are rather poorly determined observationally [the former parameter enters implicitly through the definition of $P(W_0\vert \Upsilon)$ (Fig. \ref{Ps}) while the latter enters explicitly in equation \ref{conv}]. Here we have taken $\alpha=-1.4,~\gamma=0.7$ (we define $\delta \equiv \alpha+\gamma$ as this is the only relevant parameter given the form of Eq. \ref{conv}; hence $\delta=-0.7$), and $\beta=0.3$. While somewhat uncertain, these values are relatively well constrained observationally and the sensitivity of the model to the small uncertainties in those parameters will be further investigated below. A good fit to the (extrapolated) $\partial^2 N(z=0)/\partial z \partial W_0$ data is obtained for $\xi_w=1.6$ and for an effective cross-section for \ion{Mg}{2}\ absorption around an $L^\star$ galaxy whose projected radius $R_{\rm MgII}(L^\star) \equiv \sqrt{\sigma(L^\star)/\pi}=\sqrt{{\xi_\sigma \sigma_{\rm HI}(L^\star)}/{\pi}}$ satisfies:
\begin{equation}
\begin{array}{lll} 
\displaystyle R_{\rm MgII}(L^\star)  & \displaystyle \simeq & \displaystyle 40 \left [ \frac{1}{0.05} \frac{\partial N(W_0>1{\rm \AA})}{\partial z}  \right ]^{1/2} f_L^{-1/2}\\
& \displaystyle \times & \displaystyle \left [\frac{\phi^\star}{3\times10^{-3}\,{\rm Mpc^{-3}}}  \right ]^{-1/2}\,{\rm kpc}. 
\end{array}
\label{reff}
\end{equation}
\citet[see their Fig. 18]{zw05} find that the typical (projected) \ion{H}{1}\ radius of an $L^\star$ galaxy at $z=0$, as detected down to a column density limit of $3\times 10^{19}\,{\rm cm^{-2}}$ is $\sim 35$\,kpc, i.e., of order $R_{\rm MgII}$ for $f_L=1$, indicating that $\xi_\sigma \simeq 1$. Such scales are also consistent with the size of the Milky Way disk measured at similarly low \ion{H}{1}\ column densities \citep{kal08}.

\begin{figure}
\plotone{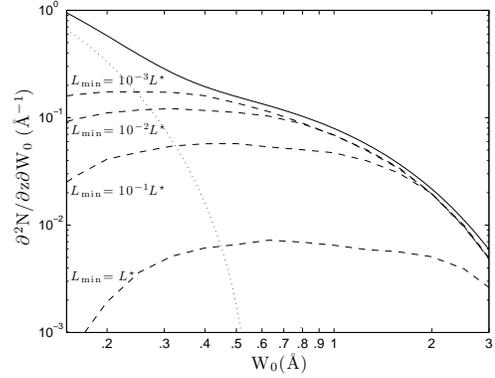}
\caption{The contribution of galaxies of different luminosities to the occurrence of \ion{Mg}{2}\ intervening systems. The contribution of all galaxy disks with luminosity $L>L_{\rm min}$ to  the $\partial^2 N(z=0)/\partial z \partial W_0$ signal is shown for several values of $L_{\rm min}$ denoted next to each dashed curve. The best-fit model of \citet{nes05} to their data (solid line) is scaled up by 20\% to allow an easy comparison with the model. Clearly, weak absorbers with $W_0<0.5$\AA\ cannot be accounted for by the model even when the contribution from very faint galaxies, with $L \ll 10^{-2}L^\star$, is taken into account. The residual signal, after the subtraction of the $L_{\rm min}=10^{-3}L^\star$ model from the data, may be interpreted as the  weak absorbers' signal contributing to the total $\partial^2 N(z=0)/\partial z \partial W_0$ distribution, and is not accounted for by our model (dotted line). The signal for strong absorbers is dominated by relatively luminous star forming, galaxies.}
\label{lmin}
\end{figure}

We examined how galaxies of different luminosities contribute to the estimated values of $\partial^2 N/\partial z \partial W_0$. We show the results in Figure \ref{lmin}.  If, as pointed out by \cite{ste94,ste97,kee06}, galaxies with $L \ll 0.1L^\star$, are seen to give rise to strong absorption, then absorbers with $W_0>0.5$\AA\ can indeed be fully accounted for by star-forming galaxy disks. Whether $L\ll 10^{-2}L^\star$ galaxies give rise to strong metal absorption systems is unclear, since our model does not require it. Furthermore, at such low luminosities, galaxy disks become \ion{H}{1}\ poor and do not seem to follow the relation assumed by equation \ref{sigma} \citep{zw05}. 

\begin{figure*}
\plotone{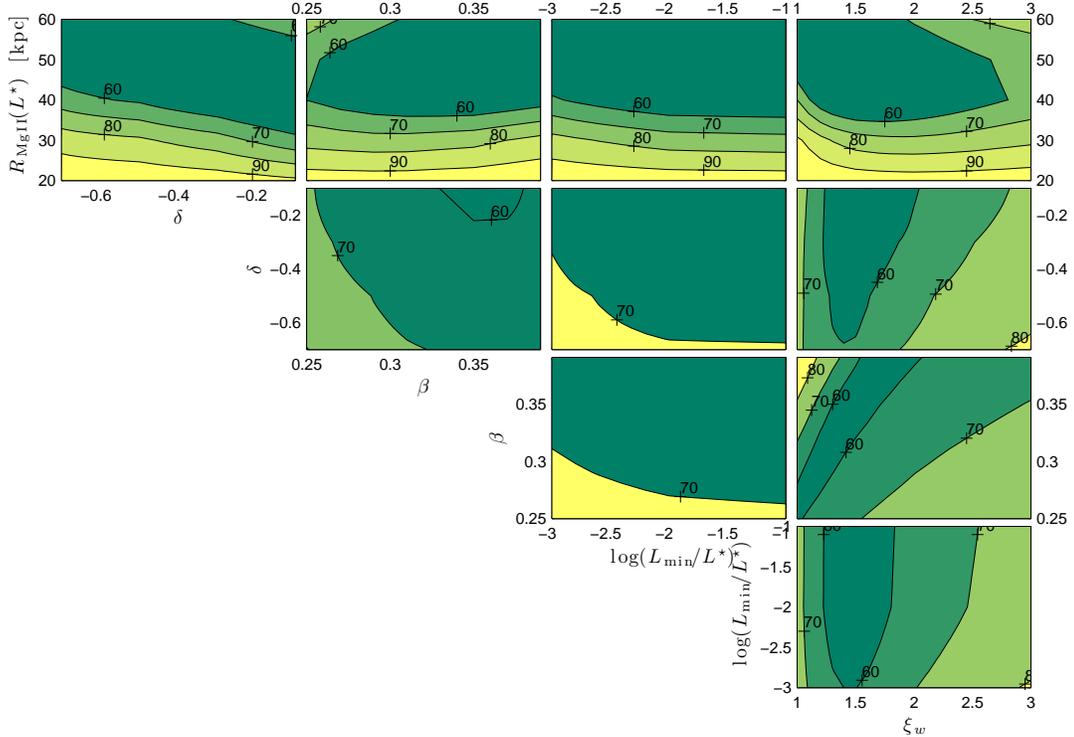}
\caption{The rejection confidence contours for our models. The multi-parameter space has been divided to subplots, each of which refers to two parameters among the five and the results for the remaining parameters have been marginalized over. Each point gives the mean rejection probability of the marginalized model, as determined from reduced $\chi^2$ statistics, $\chi_\nu^2$. The mean value of $\chi_\nu^2$ at each point (not shown) is the geometric mean of the reduced $\chi^2$ of all models being marginalized over. The $60\%$ contour level (see labels) correspond to a $\chi_\nu ^2\simeq 1$. It should be emphasized, however, that even in regions whose (mean) rejection probability is relatively high, there may be individual models providing a very good fit to the data. Clearly, model predictions are in good agreement with the data for the various parameter ranges considered here. Model predictions are particularly sensitive to the values of $R_{\rm MgII}(L^\star)$ and $\xi_w$ with the other parameters having a lesser effect on the quality of the fit.}
\label{chi2}
\end{figure*}

It is important to note that our model does not reproduce the number density of {\it weak} absorbers with $W_0\ll 0.5$\AA.  As can be seen in Figure \ref{lmin}, at these equivalent widths, there is a clear discrepancy between the number density of \ion{Mg}{2}\ lines predicted for all galaxy luminosities in our model, and the observed number density of \ion{Mg}{2}\ systems.  This difference suggests that the physical origins of weak \ion{Mg}{2}\ systems may be different from that proposed herein \citep{ri09}. This conclusion is related to and further strengthened by the fact that $W_0\sim 0.5$\AA\ is also the value at which a break is seen in the $\partial^2 N(z>0)/\partial z \partial W_0$ function \citep{nes05}.

Consider now the sensitivity of the predictions for $\partial^2 N(z>0)/\partial z \partial W_0$ to the various parameters of the model.  In particular, $\xi_\sigma$ (or $R_{\rm MgII}$ ) is essentially unknown, as is $\xi_w$. [One may obtain indirect statistical information regarding $\xi_w$ from the data of \citet{we08} yet its interpretation in terms of $\xi_w$ is not unique; see \S5.4]. We find that our model predictions change in the following way: varying $\xi_w$ changes the shape of the predicted curve such that larger $\xi_w$ results in a more shallow decline at the high-$W_0$ end. This is due to fainter and more numerous galaxies contributing to the signal at the high-$W_0$ end. Varying $\xi_\sigma$ [or $R_{\rm MgII}(L^\star)$] merely changes the normalization of the models but maintains their shape unaltered. The same behavior applies for $f_L$. In addition to the above parameters, we have also considered the effect of measurement uncertainties on the observationally deduced values used by our model and held fixed in the above analysis. For example, adopting smaller values of $\beta$ result in more low luminosity objects giving rise to stronger lines giving rise to a flatter $\partial^2 N(z>0)/\partial z \partial W_0$ distribution. Likewise, increasing $\delta$ results in a smaller contribution to the number density from low luminosity objects (since the product of their density and cross-section - i.e., their opacity - is smaller), creating a flatter $\partial^2 N(z>0)/\partial z \partial W_0$ curve for weak lines.

In order to quantify the degeneracies inherent in the model, and to constrain the physics most relevant to forming strong intervening systems in galaxy disks, we have considered a range of values for all the relevant parameters in our model: $20<R_{\rm MgII}(L^\star)<60$\,kpc ($0.3 \lesssim \xi_\sigma \lesssim 2.3$), $-0.7<\delta<-0.1,~0.25<\beta<0.4,~10^{-3}<L_{\rm min}/L^\star<0.1$, and $1<\xi_w<3$. In this multi-parameter space, we have calculated a grid of models and evaluated their  agreement with the data using $\chi^2$ statistics. Examples for a few such models, and their agreement with the data are shown in Figure \ref{dndw} (dotted curves). Presenting the regions in phase space for which the fits are acceptable, we show in Figure \ref{chi2} various contour plots as a function of two parameters at a time. The quantity plotted is the {\it mean} significance with which models may be rejected. It should be emphasized that, in each inset, we have marginalized over the remaining parameters of the model such that the geometric mean of the $\chi^2$ values has been evaluated at any point and the mean significance accordingly calculated. We also note that, due to our marginalization, it is possible, in principle, to find very good agreement between specific models and the data also in  parameter space regions where the mean significance is low. Clearly, the models are in good agreement with the data for the range of parameters discussed above. For example, models for which the size of the \ion{Mg}{2}\ disk of an $L^\star$ galaxy, is greater than $\sim 30$\,kpc (for $f_L=1$ and with the disk size of other galaxies following from eq. \ref{sigma}),  can explain the observed  $\partial^2 N/\partial z \partial W_0$  distribution, given the uncertainties. To conclude, the fit is sensitive to $\xi_\sigma$ and $\xi_w$, and to a lesser extent on  the values of $\delta$, $\beta$, and $L_{\rm min}$ within the observationally motivated range of values. This also means that our model is, effectively sensitive only to $\xi_\sigma,~\xi_w$, and $\eta$ and therefore provides a $\le 3$ parameter fit to the data.

It is evident from Fig. \ref{chi2}, that good fits are obtained for $\xi_w\gtrsim 1$. In particular, the model wherein the outflow velocity field across the disk is uniform, suggests that $1<\xi_w<2$. Therefore, our model predicts that the \ion{Mg}{2}\ transitions extend to somewhat higher velocities than the \ion{Na}{1}\ lines. This is reasonable given the higher magnesium abundance with respect to sodium (while neglecting ionizing corrections), and the stronger oscillator strengths of the lines in the former case. Alternatively, the apparently higher velocities for the \ion{Mg}{2}\ transitions could be a purely observational effect, as discussed in \citep{ma09}. Our conclusion that $\xi_w \gtrsim 1$ is further supported by the difference in the \ion{Mg}{2}\ line profiles seen towards the centers of galaxies by \citet{we08}, compared to the \ion{Na}{1} profiles observed by \citet{ma05}. We return to this point in \S5.4.

If most star-forming galaxies brighter than $10^{-2}L^\star$ give rise to strong \ion{Mg}{2}\ absorption then Figure \ref{chi2} suggests that $35<R_{\rm MgII}(L^\star)<60$\,kpc for the studied parameter space. Such sizes are comparable to those deduced by \citet[see their eq. 3]{kac08} given the occurrence of strong systems and based on a very different set of arguments.  Our deduced sizes are also consistent with the recent findings of \citet{me10} who show that the \ion{O}{2}\ luminosity function, which traces star-forming regions in the universe, can be fully accounted for by the number density of strong \ion{Mg}{2}\ systems that surround star-forming galaxies on similar scales. Interestingly, such sizes have also been found to characterize the  \ion{H}{1}\ disks of $L^\star$ galaxies, as measured for the {\it Westerbork HI Survey of Irregular and Spiral Galaxies} (WHISP) sample \citep{vdh01}. In particular, \citet[see their Fig. 18]{zw05} found a projected \ion{H}{1}\ radius of $L^\star$ galaxies to be in the range 25-50\,kpc. This implies that,  in our model, $\xi_\sigma \sim 1$. This conclusion is consistent with the findings of \citet[see their Fig. 29]{bo95} who showed that the projected radius of the \ion{H}{1}\ disk for an $L^\star$ galaxy is $\gtrsim 25$\,kpc, and that the size of the \ion{Mg}{2}\ absorption cross-section is of at least a comparable size. This means that the cross-section for strong \ion{Mg}{2}\ absorption is much larger than that of the optical and UV emitting disks. This makes outflows visible for studies in which the galaxy itself acts as a background source against which \ion{Mg}{2}\ systems that are associated with the galaxy are detectable \citep{tr07,we08}. 

The match between \ion{H}{1}\ disk sizes and the predicted $R_{\rm MgII}$ is also consistent with properties of the \citet{rao06} sample of absorbers for which the typical \ion{H}{1}\ column densities that are associated with $W_0\gtrsim 0.5$\AA\ systems are of order $\gtrsim 10^{19}\,{\rm cm^{-2}}$. Such \ion{H}{1}\ column densities are somewhat below the limiting column densities of galaxies in the WHISP sample, which is $\sim 3\times 10^{19}\,{\rm cm^{-2}}$ \citep{zw05}. Given that the apparent disk sizes increase with increasing sensitivity of the observations (so that lower gas columns can be probed), and extrapolating the results of \citet[see their Fig. 18]{zw05}, we find that the projected radii of $L^\star$ galaxies down to a column density of $10^{19}\,{\rm cm^{-2}}$ is $\gtrsim 40$\,kpc, in line with our findings. We note that about 50\%, of \ion{Mg}{2}\ systems with $0.5<W_0<1$\AA\ have \ion{H}{1}\ column densities smaller than $5\times 10^{18}\,{\rm cm^{-2}}$ \citep{rao06} i.e., considerably below the detection limit of the WHISP sample. This implies that the cross-section for dilute low column density gas, as appropriate for $W_0\gtrsim 0.5$\AA\ systems, is essentially unknown and may be larger than that which is currently deduced from low sensitivity 21\,cm emission studies [see e.g., \citet{bra04} who find evidence for an extended envelope of low column density gas on scales $>50$\,kpc around M31]. 

Further evidence that strong \ion{Mg}{2}\ systems are probing disks comes from the fact that the coherence size of $W_0\sim 0.5$ absorbers, as estimated from observations of lensed quasars, is found to be $3^{+3}_{-1}$\,kpc \citep{el04}. In our model, such systems predominantly arise in low luminosity galaxies with $10^{-3}L^\star<L<10^{-2}L^\star$  (see Fig. \ref{lmin} and below). As such, our model implies a coherence length of order the disk radius, i.e., $R_{\rm MgII} (L/L^\star)^{\gamma/2}\sim 3.5-8$\,kpc (for $f_L=1$), a value that is consistent with observations. 

\subsection{The Kinematic Properties of Outflows Across Disks}

\begin{figure}
\plotone{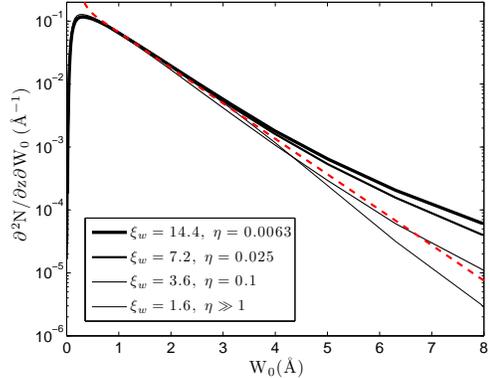}
\caption{Predictions for $\partial^2N/\partial z \partial W_0$ for several values of $\xi_w$ and $\eta$: models with $\xi_w\sim \eta^{-0.5}$ and with all other parameters held fixed are in good agreement with the data of \citet[dashed red curve]{nes05} in the range $0.5<W_0<4$\AA. Nevertheless, for larger $W_0$, predictions are sensitive to the value of $v_w$. In particular, larger $v_w$, and correspondingly smaller $\eta$, result in a powerlaw decrease rather than an exponential decay in the number of large $W_0$ systems. This results from the number statistics of such systems being dominated by relatively faint star-forming galaxies for which quasar sightlines pass close to their centers rather than the exponential tail of the galaxy luminosity function. Note the sharp rise of the observed  $\partial^2N/\partial z \partial W_0$ for $W_0\lesssim 0.5$\AA\ resulting from the unique population of weak systems, which is not accounted for by the model.}
\label{gal_kin}
\end{figure}

In \S3.1 we defined the projected core radius of the disk, within which the outflow velocity approaches a constant value, to be $\rho_0(L)\equiv \eta R_{\rm MgII}(L)$. In our model for the kinematics of outflows across disks (Eq. \ref{vwb}), we adopt $\eta=0.1$ for no other reason than the fact that, in this case, $\rho_0$ coincides with the optical, star-forming size of the disk.  Running Monte Carlo simulations and including the effect of random sight lines through the disk surface, we find that it is possible to fit the data well if $\xi_w=3.6$. This means that, for an $L^\star$ galaxy,  the core in the velocity profile occurs for $\rho_0(L^\star) = \eta R_{\rm MgII}(L^\star) \sim 4$\,kpc, while the maximum velocity of the \ion{Mg}{2}\ lines is $\gtrsim 3$ times higher than that measured for the \ion{Na}{1}\ lines. For galaxies with $\Upsilon \sim 10\,{\rm M_\odot~yr^{-1}}$, this model predicts maximum wind velocities of $\gtrsim 250 \times 10^{0.3} \gtrsim 500\,{\rm km~s^{-1}}$ toward their cores. Note, however, that the mean velocity (and conversely $W_0$), as measured for a statistical sample of such galaxies, will be lower since only a fraction of all objects would launch outflows from their centers that travel at the maximum velocity according to $P(W_0 \vert \Upsilon)$ (see also \S5.4). 

The above solution is, however, non-unique, and equally good fits may be obtained, for example, when taking $\xi_w=6$ and $\eta=0.03$ (while holding all other parameters fixed). In particular, the current data quality for $\partial^2 N/\partial z \partial W_0$, as extrapolated to $z\sim 0$, and for large values of $W_0$, provides little constraints on the allowed $\xi_w,~\eta$ ranges, and that $\xi_w \sim \eta^{-0.5}$ traces a good solution curve in phase space with all other parameters held fixed and given the current data; see figure \ref{gal_kin}. However, different models have different predictions for the number statistics of the strongest systems. In particular, in models for which $v_w$ is higher (and $\eta$ correspondingly smaller), the abundance of the strongest systems is dominated by sightlines passing close to the centers of relatively faint galaxies (see above). Currently, the $\partial^2 N/\partial z \partial W_0$ data set alone cannot be used to reliably constrain physical models for galactic outflows. As we discuss in \S5.4, however, there are additional data that may be used to shed light on the kinematics of galactic outflows.

\section{Discussion}

A Kennicutt-Schmidt law for strong absorbers is consistent with the notion that most strong \ion{Mg}{2}\ systems seen in the spectra of background quasars arise in disks of intervening star-forming galaxies.  We have shown that the density distribution of strong \ion{Mg}{2}\ absorbers, $\partial^2 N(z=0)/\partial z\partial W_0$, can be fully accounted for if galaxy disks with the same properties as those observed in low-redshift galaxies are responsible for the absorption. Our model implies that (a) the \ion{Mg}{2}\ absorption cross-section for a galaxy is comparable to that of the \ion{H}{1}\ cross-section of nearby galaxy disks, and that (b) the velocity of the outflowing gas that gives rise to intervening  \ion{Mg}{2}\ absorption is consistent, up to a factor of order unity, with the outflow kinematics measured for \ion{Na}{1}\ and \ion{Mg}{2}\ absorption troughs towards the centers of local star-forming galaxies.  The notion that strong \ion{Mg}{2}\ systems are material in and outflowing from galaxy disks is also consistent with the recent findings of \citet{ber08} indicating that the magnetic fields associated with strong systems are comparable to galactic values.

The association of strong \ion{Mg}{2}\ systems with galaxy disks is also physically appealing as it naturally resolves several outstanding theoretical problems in the field: for example, if strong \ion{Mg}{2}\ systems are interstellar material in extended  disks then the problem of cloud formation, confinement, and stability in dilute intergalactic environments \citep{bi09} is, clearly, alleviated. Furthermore, there is no need to transport cool metal-rich material from galaxy centers to large scales while avoiding over-ionization and mixing. In fact, in our model, the physics of strong intervening \ion{Mg}{2}\ systems is exactly that of the ISM, which is (at least initially) gravitationally bound and enriched by stars in its environment. 

Below, we discuss a few implications of our results.

\subsection{Absorber-Galaxy Associations}

\begin{figure*}
\plotone{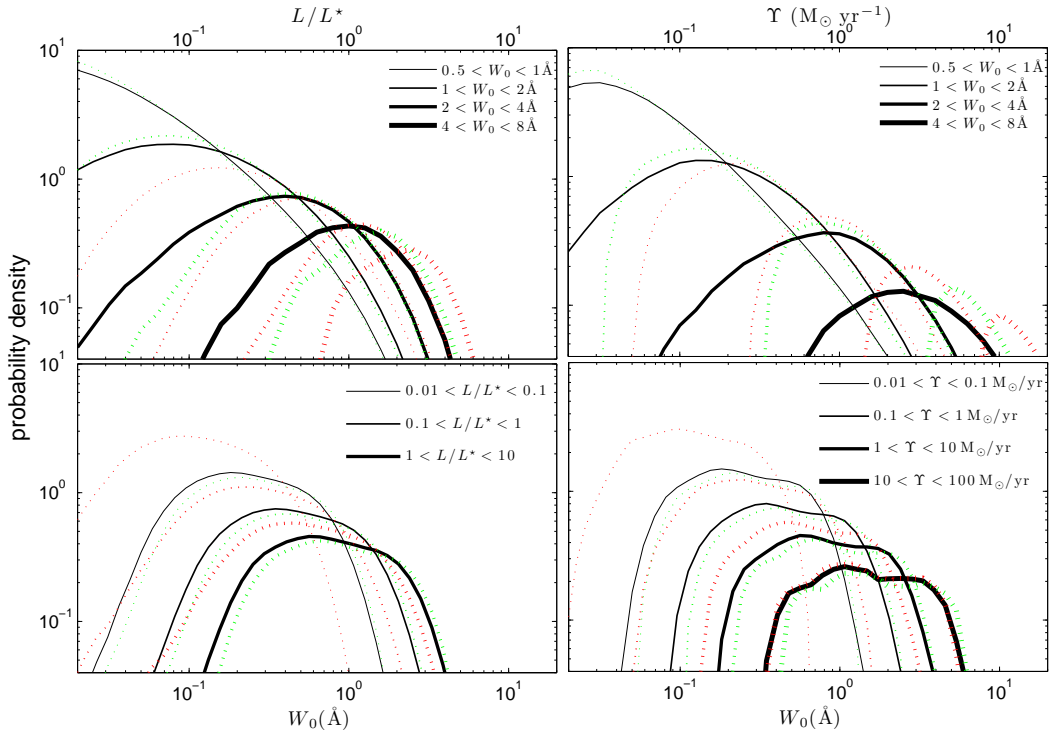}
\caption{{\it Upper panels}: The predicted probability distributions for the luminosity (left) and star formation rate (right) for absorber-selected host galaxies. The data are shown for different $W_0$ bins (see legend) and for the three models considered in this work (model C in red, model B in black, and model A in green). Clearly, a broad range of galaxy luminosities and star formation rates is selected by absorbers over a narrow $W_0$ range, which could explain the lack of significant observational correlations between their properties. Generally, however, stronger systems tend to pick more luminous and more star forming galaxies. There is a qualitative agreement between the two models, with model B being shifted to somewhat lower galaxy luminosities and star formation rates (see text). {\it Lower panels}: Model predictions for the $W_0$-distributions as would be observed for galaxy-selected absorbers as a function of the galaxy luminosity (left) and star formation rate (right; note that values in the legend are in units of solar masses per year). Here too, the distributions are broad with an overall tendency for brighter, more star forming galaxies, to give rise to stronger systems.}
\label{dist}
\end{figure*}

As demonstrated above, a broad range of galaxy luminosities gives rise to strong \ion{Mg}{2}\ absorption. For $W_0\gtrsim 0.5$\AA\ systems, the galaxies which give rise to the absorption have luminosities $>10^{-2}L^\star$. For example, the model shown in Figure \ref{lmin} predicts that there is a 50\% chance that the hosts of $W_0\simeq 0.7$\AA\ \ion{Mg}{2}\ systems are fainter than $0.1L^\star$. It may therefore be challenging to identify the galaxy which directly hosts the absorber, since it could be outshone by the nearby quasar. In such cases only bright hosts or neighboring galaxies that are clustered around the true host may be detected (provided they are bright enough and far enough from the quasar sight line).  

Given the predictions of our model, a better approach to studying the origin of $W_0\lesssim 1$\AA\ systems at high redshifts is to use $\gamma$-ray bursts (GRBs) as background sources rather than quasars [e.g., \citet{ver09} and references therein]. With this technique, deep images of the fields obtained after the GRB light has faded is more likely to reveal fainter galaxies, even when the GRB host galaxy is also present \citep{pr07,th08,che09}.

Alternatively, a comprehensive study of absorption lines arising from very low redshift galaxies could be used to test the predictions of our model. Although HST is required to record spectra of UV lines at $z\sim 0$, nearby galaxies probed by quasar sightlines can be selected a priori by the same properties studied in our model, such as their luminosity, morphology, star formation rate, etc.  Probing a significant number of galaxies with a range of these properties would enable us to map the distribution of metal line properties directly. The installation of COS aboard HST, with a sensitivity higher than the spectrographs previously available on the satellite, now makes a large survey possible.

Our model gives specific predictions for the statistical relations between the absorber properties (namely, $W_0$) and its host galaxy properties (such as $\Upsilon$ and $L$). Nevertheless, judging from Figure \ref{chi2}, the allowed parameter space in which there is good agreement between our model and the available \ion{Mg}{2}\ data, is large. In this section we therefore focus on a small subset of all models whose parameterizations are given in table \ref{tab}, and all provide equally good fits to the data (specifically, all models trace the dashed black curve in Fig. \ref{dndw}). This set of models which we label A, B, and C, is consistent with the notion that galaxies much fainter than $0.01L^\star$ contribute little to the cross-section for $W_0\gtrsim 1$\AA\ systems. This agrees with recent findings that galaxies much fainter than $\sim 0.01L^\star$ are gas deficient and contribute little to the matter density in  \ion{H}{1}\ \citep{zw05}. Figure \ref{dist} shows the predicted luminosity, star formation rate, and rest-equivalent width distributions: upper panels are relevant to situations where galaxies are selected based on their associated absorber properties, and the lower panels to cases in which a galaxy sample is first obtained and follow up observations are then carried out to look for their associated absorption systems. As shown, weak absorbers that arise in disks select for fainter star forming hosts  while strong absorbers tend to be associated with  brighter star-forming objects. Nevertheless, our model does not preclude the possibility that some strong systems arise in early-type galaxies; statistically, however, the number counts of strong systems can be fully accounted for by star-forming galaxies alone. 

More quantitatively, our models suggest that the typical luminosities of the hosts of  $W_0\sim 2$\AA\ systems are around $0.1\,L^\star$. This seems to be consistent with the recent findings of  \citet{ch09} who found  faint $L\lesssim 0.1L^\star$ galaxies in the fields of such absorbers. According to the model, by selecting the rare, $W_0>4$\AA\ absorbers, one is most likely to select galaxies with $L \gtrsim L^\star$ and with a star formation rate $\Upsilon>1\,{\rm M_\odot\,yr^{-1}}$. Nevertheless, the exact numbers depend on the particular model chosen, as can be seen by comparing models A, B, and C.  For example, models B and C give very different predictions for the star-formation rate distribution of $W_0>4$\AA\ hosts such that, for model C, a considerable fraction of hosts have $\Upsilon>7\,{\rm M_\odot~yr^{-1}}$ while, for model B, galaxies with $\Upsilon\gtrsim 1\,{\rm M_\odot~yr^{-1}}$ can substantially contribute to the number of such systems. The different predictions in each case imply that reliable measurements of the $L(W_0)$ distributions will constrain the model.  Further, all models predict that galaxies with luminosities $L^\star \lesssim L\lesssim 6L^\star$ give rise to $W_0>4$\AA\ absorbers. The fields of such ultra-strong systems have recently been imaged by \citet{nes07} who find an excess  of $L>L^\star$ near quasar sightlines. This fact is consistent with our findings, although a more quantitative comparison awaits secure identifications of galaxies in those fields.

Our model implies that surveys which first select host galaxies according to their luminosity and star formation rates, and then study absorption through their disks, would see a broad range of absorber properties at any given luminosity and SFR bin. For example, a faint dwarf with $L\sim 0.01L^\star$ is likely to give rise to an absorber with $W_0\sim 0.6$\AA. However, the chances of a very bright galaxy, $L\gtrsim 1L^\star$ giving rise to the same $W_0$ are also considerable (see Fig. \ref{dist}). This may be the reason for the scarcity of clear relations between galaxy and absorber properties (in addition to uncertainties related to the identification of the true host). The model is qualitatively consistent with the recent findings of \citet{bou07l}, \citet{wi07}, and \citet{me10}, and which are further discussed in \S5.6.  Our results highlight the need for large samples of quasar-galaxy pairs, which would overcome the large scatter in galaxy-absorber properties and help to reveal the underlying galaxy-absorber association.  

The findings presented here are consistent with the notion that most large-scale galaxy-absorber associations, where the impact parameter is much larger than the \ion{H}{1}\ disk size of the true host galaxy, arise from galaxy-galaxy clustering. This agrees with recent findings concerning the low covering factor (of order 20-50\% within a few tens of kpc of the host galaxy) for \ion{Mg}{2}\ absorption on such scales \citep[and also Bowen \& Chelouche 2010]{tr05,ch08,ba09,gau10}. This may also be the reason for some edge-on galaxies apparently being associated with strong absorption along sightlines at right-angles and far above their disks \citep{ste02}. It is possible, in principle, to use our model and some knowledge of galaxy---galaxy clustering to predict the $W_0$ distribution and covering factor as a function of the impact parameter from bright galaxies, which are not necessarily the true hosts of the absorber, and to compare to the data \citep[and also Bowen \& Chelouche 2010]{ch08}. This raises the interesting possibility that galaxy-galaxy clustering and group kinematics may be directly probed by absorber-galaxy studies leading to a better understanding of structure formation and to additional constraints on halo occupation distribution (HOD) models \citep{bou06,zz07,ti08, lun09}. 

According to the model, absorption through our own Galaxy disk ($L\sim L^\star$ and $\Upsilon \sim 1\,{\rm M_\odot\,yr^{-1}}$) would yield $W_0\sim 1.2\pm0.8$\,\AA\ (this was calculated from the relevant $W_0$ distributions with the uncertainty corresponding to one standard deviation around the average). The mean $W_0$ or outflow speed from the Galactic disk, as would be measured from \ion{Mg}{2}\ absorption lines toward extragalactic objects, is $\simeq W_0/2\sim 0.6\pm0.4$\AA\ or $\sim 60\pm40\,{\rm km~s^{-1}}$ (neglecting the possibly important contribution of galactic rotation to the observed profile along some sight lines to extragalactic objects). This is consistent with the results of \citet[see their Fig. 1]{bo95b} who found similar velocity dispersion in several local \ion{Mg}{2}\ systems toward quasars. We note that our analysis predicts that the observed velocities of the \ion{Na}{1}\ lines should be a factor $\gtrsim 2$ lower than those measured for \ion{Mg}{2}\ (see Table \ref{tab}); i.e., of order $30\pm15\,{\rm km~s^{-1}}$ as would be observed along extragalactic sight lines. This value is qualitatively consistent with that observed for the \ion{Ca}{2}\ lines  \citep[see their Fig. 5]{bo91}. A more quantitative comparison between the kinematics of different ions requires better understanding of the structure and dynamics of galactic winds.

Lastly, we note that our prediction for the Galactic wind velocity is similar to the outflow speed predicted at one pressure scale-height above the disk by \citet{ev08} based on fits to the large-scale diffuse X-ray emission and using an energy driven outflow model. It is therefore possible that the outflowing cool material which is probed by \ion{Mg}{2}\ lines provides the necessary initial conditions for the onset of a large scale thermal wind, as advocated by these authors (see also \S5.9). Therefore, even if intervening \ion{Mg}{2}\ systems relate only to outflowing matter within one pressure scale-height above the disk, they could be related to large scale outflows that ultimately escape the galaxy and perhaps join the IGM; we return to this point in \S5.9.

\begin{table}
\caption{Model parameterization}
\begin{tabular}{llllllll}
\tableline
		&	$L_{\rm min}$	& 		& 		& $R_{{\rm MgII}}(L^\star)$ &  		&  		&\\
model	&	($L^\star$) 	& $\delta$	& $\xi_w$	& (kpc) 				  & $\eta$ & $\beta$ &   color coding \\
\tableline
A & 0.001 & -0.7 & 1.6 & 40 & $\gg 1$ & 0.3 & green\\
B & 0.01 & -0.7 & 3.6 & 40 & 0.1 & 0.3 & black \\
C & 0.01 & -0.2 & 2 & 60 & $\gg 1$ & 0.42 & red \\
\tableline
\end{tabular}
\label{tab}

\end{table}

\subsection{Metallicity and Dust Content}

\begin{figure}
\plotone{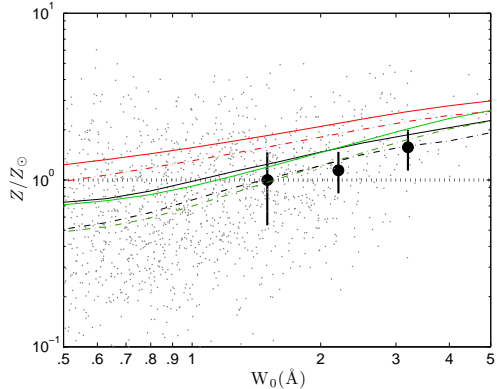}
\caption{The $W_0-Z$ relation for strong absorbers at $z=0$ as predicted by models A (green lines), B (black curves) and C (red lines). For each model the mean (solid lines) and median (dashed lines) are  shown. In addition, the typical scatter in the predicted data is shown in gray points (calculated for model B). Although a large scatter is seen, there is a clear trend between the mean or median metallicity and $W_0$. Overlaid are data points for strong intervening systems for which the mean metallicity in each bin has been estimated from the mean dust to gas ratio; see \citet{me09}. Overall, our models are in good agreement with the data of \citet{me09}, as extrapolated to low redshifts. Quantitatively, however, model predictions rely on the metallicity-luminosity relation whose whose applicability to absorption systems is not well demonstrated. Furthermore, better direct estimates for the metallicity of a large sample of strong intervening systems at low redshifts are required make a proper comparison with model predictions.}
\label{metal}
\end{figure}

If strong \ion{Mg}{2}\ systems arise from the ISM within, and outflowing from, galactic disks, then our model can be used to predict the metallicity, $Z$, of strong absorbers, as would be observed in the {\it local} universe, using the metallicity-luminosity relation for $z=0$ galaxies. In particular, we take a log-normal distribution whose mean luminosity-dependent metallicity is $\simeq 1.5Z_\odot (L/L^\star)^{0.5}$ with a scatter of order $\pm0.2$\,dex around the mean ($Z_\odot$ is the solar metallicity). This is consistent with the results of \citet[see however \citet{kew08} for uncertainties in this relation]{tr04}. This defines the probability $P(Z \vert L)$. The metallicity distribution for an absorber of a given $W_0$ is, by construction of our model, $P(Z \vert W_0)=\int dL P(Z \vert L) P(L \vert W_0)$. Using these distributions, we have carried out Monte Carlo simulations and constructed a mock sample of absorbers  and their associated galaxies, which is statistically identical to that used in the previous sections yet now includes also the metallicity for each system.  The results are shown in Figure \ref{metal} where  a relation between the mean (or median) metallicity and $W_0$ is well defined. However, the scatter is large and individual systems having the same $W_0$ may have their metallicities differ by an order of magnitude.  In particular, strong systems with $W_0 \gtrsim 1$\AA\ have mean metallicities of $\left < Z \right > \gtrsim Z_\odot$ (e.g., $W_0\sim 5$\AA\ systems have $\left < Z \right > \gtrsim 2 Z_\odot$) while weaker systems, which originate in disks, have sub-solar metallicities. Evidently, the exact numbers depend on the particular model chosen. We emphasize that these results are model predictions for the metallicity of strong \ion{Mg}{2}\ systems, as would be observed at $z=0$. Predictions for higher redshift systems will require knowledge of the metallicity of high redshift disks. A quantitative extrapolation of our results to high redshifts is beyond the scope of the present work.  

Comparing to data in the literature, we assume that the dust-to-gas ratio of strong intervening systems can be used as a proxy for their metallicity. We therefore take the statistically measured dust-to-\ion{H}{1}\ column ratio, as converted to metallicity and extrapolated to $z=0$ by \citet{me09}, and plot the mean ratio and its uncertainty in Figure \ref{metal}. As can be seen in the figure, both the mean and  median metallicities derived using model C predict values that are larger than observed. Models A and B, however, agree with current measurements surprisingly well. In particular, the predicted trend whereby weaker systems have a lower mean metallicity is in nice agreement with the observations, and the predicted values are consistent (up to $\sim 20$\% for the mean) with the data within the quoted uncertainties. The obtained $W_0-\left < Z \right >$ relation may also explain the relatively rapid increase in the equivalent widths of \ion{Fe}{2}\,$\lambda 2600$ absorption line with $W_0$ [at a given \ion{H}{1}\ column; see Fig. 1 in \cite{me09}] since stronger systems are more metal-rich. 

The agreement between model predictions and the extrapolated data is surprising given the crude metallicity estimators used. For example, the metallicity-luminosity relation of \citet{tr04} was determined from metallicity measurements of  \ion{H}{2}\ regions on compact scales in galaxies. However, it is not clear whether the metallicities of gas in galactic \ion{H}{2}\ regions need be the same as interstellar gas between \ion{H}{2}\ regions. In this respect, it is interesting to note that \citet{bo05} found good agreement between the metallicity of a Damped Ly$\alpha$ absorption (DLA) system [absorption by neutral hydrogen with column densities $\log N_{\rm HI}> 20.3$ \citep{wol86}] measured from its metal absorption spectrum and that deduced from emission. There are, however, contrary examples where \ion{H}{2}\ region metallicities do not agree with interstellar absorption metallicities \citep{can05,ch05}, although untangling the effects of metallicity radial profiles \citep{uri02,br04} --- and indeed, deciding whether the correct galaxy has been identified as the origin of the DLA system --- complicates the task of matching emission and absorption metallicities. Our models do not include metallicity gradients, an omission that will lower the predicted metallicities for sight lines that intercept the outer regions of disks. While metallicity gradients can have a crucial effect on the large-scale extinction through extended galaxy disks \citep{za94,hol05,c07,me10b} and may be easily incorporated into our calculation scheme, we believe that more elaborate models are currently unwarranted as the involved uncertainties on the metallicity of disks on large scales are large \citep{uri02,boi04}. Better estimations of the metallicity or dust-to-gas ratio of $W_0 \gtrsim 0.5$\AA\ systems promise to shed light on the metallicity gradients in the outer regions of disks.

DLAs are known to consistently show metallicities less than the solar value [e.g., \citet{pet00,nes03,kha04,che05,per07}]. While at first this seems to contradict our model predictions, we note that the probability for obtaining a DLA for a strong intervening system with a given $W_0$ is similar, within a narrow $W_0$ range, for all systems with $W_0\geq 0.6$\AA\ \citep{rao06} with no DLAs currently selected by $W_0<0.6$\AA\ systems. Therefore, most DLA surveys, which select by \ion{H}{1}\ column density may, in fact, select $0.6<W_0<1.5$\AA\ systems being the most abundant (see Fig. \ref{dndw}). The metallicity associated with such systems, in our model, is indeed sub-solar. Furthermore, DLA systems are observed at high redshifts while our formalism was developed (and applied for extrapolated \ion{Mg}{2}\ data) at $z\simeq 0$. If the evolution in metallicity of DLA systems is similar to that of \ion{Mg}{2}\ systems then we expect their typical metallicity to be lower still. Specifically, using the evolution in the dust-to-gas ratio from \citet[see their Fig. 3]{me08} as a proxy to the evolution in the metallicity, one finds that typical DLA metallicity should be $<0.2Z_\odot$ at $z\gtrsim 1.5$. At still higher redshifts $z\sim 3$, and using the metallicity evolution curves of \citet{ku02}, the typical metallicity is expected to be $<0.02Z_\odot$, and in agreement with observations \citep{ku05}. We note, however, that the scatter around the mean value is expected to be large (see Fig. \ref{metal}), which could account for the large range of metallicities reported by \citet{per07,per08}, \citet{fr10}, and \citet{ka10}.  In addition, extinction effects could bias magnitude-limited samples against finding high metallicity, high dust content DLA systems \citep[and references therein]{bou08}. 

\subsection{The \ion{H}{1}\ distribution of strong \ion{Mg}{2} Systems}

\begin{figure}
\plotone{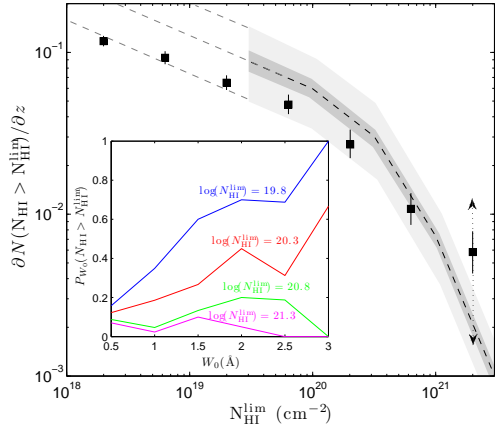}
\caption{The number density of \ion{Mg}{2}\ systems per unit redshift, $\partial N(z=0)/\partial z$ with a \ion{H}{1}\ column density above a threshold, $N_{\rm HI}^{\rm lim}$, as expected for strong systems with $W_0>0.5$\AA\ (black squares). This was calculated using $\int_{0.5{\rm \AA}}^\infty dW_0(\partial ^2 N/\partial z\partial W_0)P_{W_0}(N_{\rm HI}>N_{\rm HI}^{\rm lim}$ where  $P_{W_0}(N_{\rm HI}>N_{\rm HI}^{\rm lim})$ is the probability for an \ion{Mg}{2}\ absorber with rest equivalent width $W_0$ to have a \ion{H}{1}\ column density greater than $N_{\rm HI}^{\rm lim}$.  This  probability was estimated from the data of \citet[see their Fig. 2]{rao06} and is shown in the inset. Also shown in dashed line are the  $\partial N(z=0)/\partial z$ as measured for nearby galaxy disks by \citet{zw05}. The uncertainty was taken to reflect potentially important systematic effects discussed in \citet{rw03} and shown as a gray area. Evidently, galaxy disks at $z=0$ can fully account for the (expected) number of $z=0$ \ion{Mg}{2}\ systems up to the highest column density bin.  However, the number counts of strong systems with the highest columns, $N_{\rm HI}>10^{21.3}\,{\rm cm^{-2}}$, is expected to be a factor $\sim 2-3$ higher than estimated based on 21\,cm emission line observation of galaxy disks (see text). That said, these estimations are sensitive to beam size effects as well as extinction bias and evolution. A crude correction for the latter effects may go either way, as shown by the arrows (potentially important correction factors for the lower column points are not shown). Error bars reflect the uncertainty on the predicted absorber signal due to Poisson errors in determining $P(N_{\rm HI}>N_{\rm HI}^{\rm lim}; W_0)$ from \citet{rao06} data set and does not include additional uncertainties related to the form of  $\partial ^2 N(z=0)/\partial z\partial W_0$.  }
\label{HI}
\end{figure}

If disk galaxies give rise to \ion{Mg}{2}\ absorption, then the observed $N_{\rm HI}$ distribution of strong systems  should be accounted for by disks.  At present, our knowledge of the \ion{H}{1}\ column densities associated with strong \ion{Mg}{2}\ systems is limited to high redshifts \citep{rao06}, while our knowledge of the \ion{H}{1}\ distribution in disks is limited to nearby objects. Here we make use of the probability for an absorber of a given $W_0$ to show a \ion{H}{1}\ column density greater than some value, $P_{W_0}(N_{\rm HI} > N_{\rm HI}^{\rm lim})$. This probability was determined from the observations for $z\sim 1$ systems by \citet[see their Fig. 2]{rao06} and is shown in the inset of Figure \ref{HI}\footnote{We note that while \citet{rao06} use various selection criteria to construct their sample, which particular set of criteria used is not important and results in a similar $P_{W_0} (N_{\rm HI}>N_{\rm HI}^{\rm lim})$ given the uncertainties.}. 

Here we {\it assume} that the above determined $P_{W_0} (N_{\rm HI}>N_{\rm HI}^{\rm lim})$ is applicable to low-$z$ systems and calculate the number density of \ion{Mg}{2}\ systems per unit redshift that give rise to $N_{\rm HI}>N_{\rm HI}^{\rm lim}$, $\partial N(N_{\rm HI}>N_{\rm HI}^{\rm lim})/\partial z=\int_{0.5{\rm \AA}}^\infty dW_0(\partial ^2 N/\partial z\partial W_0)P_{W_0}(N_{\rm HI}>N_{\rm HI}^{\rm lim})$. We take $\partial ^2 N/\partial z\partial W_0$ from our best-fit model to the extrapolated data of \citep{nes05} and show the results in Figure \ref{HI}. The uncertainties on the predicted $\partial N(N_{\rm HI}>N_{\rm HI}^{\rm lim})/\partial z$ signal shown in Figure \ref{HI} include only those due to Poisson noise in determining  $P_{W_0} (N_{\rm HI}>N_{\rm HI}^{\rm lim})$ from the \citet{rao06} data and do not include additional uncertainties of a factor $\gtrsim 3$ on the amplitude of the $\partial^2N/\partial z \partial W_0$ signal (see \S3). Comparison with the \ion{H}{1}\ data of \citet[see their Fig. 8]{zw05}, which were obtained from 21\,cm emission observations of local galaxy disks shows overall agreement: the shape of the $\partial N(N_{\rm HI}>N_{\rm HI}^{\rm lim})/\partial z$ curve is reproduced given the uncertainties. Furthermore, the amplitude is also consistent within the uncertainties lending further credibility to the model and our estimates for the cross-section of galaxy disks. The uncertainties on the 21\,cm emission data show both the statistical errors reported by \citet{zw05} as well as possible systematic effects discussed in \citet{rw03}. For $N_{\rm HI}^{\rm lim}<10^{19.8}\,{\rm cm^{-2}}$, the WHISP sample suffer from sensitivity issues and is therefore incomplete. However, an extrapolation of the powerlaw trend at low column densities as implied by 21\,cm observations is seen to be consistent with the much lower columns probed by strong \ion{Mg}{2}\ systems. This may be related to the fact that strong systems seem to trace the Galactic dust-to-gas relation \citep{bo78,me09}.

As is evident from Figure \ref{HI}, the number counts of strong systems with the highest $N_{\rm HI}$ seem to be higher by a factor of $\sim 3$ than implied by the 21\,cm emission data. This fact was already noted by \citet{zw05} and it remains to be seen whether it is persists when better statistics are gathered with COS. The reason for this mismatch is not known and several possible factors that could cause it: for example, if the local Schmidt-Kennicutt law holds also at high redshift then the fact that the global star formation rate increases with redshift \citep{lil96,mad98}, may mean that $N_{\rm HI}$ increases with redshift, possibly in a complicated manner. Therefore, it may be that our assumption that $P_{W_0} (N_{\rm HI}>N_{\rm HI}^{\rm lim})$, as derived for high-$z$ systems, is directly applicable to the local universe is unwarranted. Specifically, given the fact that the global star formation rate has decreased by about an order of magnitude from $z\sim 1$ to present times, and using the Schmidt-Kennicutt scaling $ N_{\rm HI} \propto \Upsilon^{2/3}$, $N_{\rm HI}$ may be lower by a factor $\sim 4$ compared to that measured by \citet[see Fig. \ref{HI}]{rao06} for high-$z$ systems. A low-$z$ analog to the \citet{rao06} compilation or higher-$z$  21\,cm observations of galaxy disks will be most useful in settling this issue. A further systematic effect may be related to the typical beam size used by each approach: in absorption line studies the quasar emission region sets the angular resolution, which is of order $\sim 10^{-8}$\,rad in diameter. In contrast, radio surveys have typical beam sizes which are $\sim 10^3$ times larger. As demonstrated by \citet[and references therein]{zw05}, lower resolution results in an under-estimation (smearing) of high column density regions. The degree to which this explains the apparent mismatch in the number of systems with the highest \ion{H}{1}\ columns is unclear and depends on the structure of the ISM. Lastly, it may also be that the highest $N_{\rm HI}$ systems are over-represented in the \citet{rao06} sample due to the combination of selection criteria used; an effect which cannot be easily identified given the small number statistics for $N_{\rm HI}>10^{21}\,{\rm cm^{-2}}$ systems in their work. We note that a recent census of DLA systems by \citet{not090} at $z\gtrsim 2$ shows consistency, up to a scaling factor, with the \citet{zw05} results.

It is worth noting that there is an additional, potentially important, systematic effect that relates to dust extinction and reddening: \citet{me08} have shown that high-column systems may be missed by magnitude-limited samples, such as the {\it Sloan Digital Sky Survey} (SDSS). Using the mean dust-to-gas ratio for strong systems from \citet{me09} and taking an SMC reddening curve, we estimate $E(B-V)\simeq 0.2$\,mag for $N_{\rm HI} \sim 10^{21.3}\,{\rm cm^{-2}}$ systems, implying a mere $\sim 50$\% completeness of the SDSS for such systems \citep[see their Fig. 2]{me08}. Higher column density regions, assuming the same mixture of dust and gas, may be completely missed by magnitude-limited samples, leading to further uncertainties concerning the true column density distribution of strong systems. It is important to try to map the outer regions of galaxies in 21\,cm down to the lowest possible columns \citep{bra04} and test whether the cross-section and column density distribution of disks and their outflows at low-$z$ can indeed explain the properties of strong systems.

Our finding that most strong \ion{Mg}{2}\ systems are accountable for by disks implies that one does not lose most strong absorption systems due to an extinction bias, as concluded by \citet{el04b}\footnote{However, one may still lose a large fraction or all of the less numerous ultra-strong systems or those fewer systems with large \ion{H}{1}\ columns; see \S5.2 and also \citet{me08}.}. In our model, the DLA and strong \ion{Mg}{2}\ systems phenomena are consistent with being two facets of the same phenomenon: ISM in (and outflowing from) galaxy disks. This is in line with the results of other studies that compared the properties of \ion{H}{1}\ and \ion{Mg}{2}\ selected DLAs finding no statistically significant difference between the two populations \citep{per04}. It is also in agreement with the cosmic mass density of \ion{Mg}{2}\ selected DLAs and \ion{H}{1}\ selected DLAs being similar [see e.g., Fig. 17 of \citet{not090}].  Our findings are also consistent with the notion that  the high-end tail of the observed \ion{H}{1}\ column density distribution, for a given $W_0$, arises in less luminous, hence metal- and dust-poor disks \citep{bou08} while sight lines through metal-rich star forming galaxies having, on average, large \ion{H}{1}\ columns and dust-rich gas, may be missed due to extinction. If true then, on average, galaxy disks allow through a considerable fraction of the the light from background sources. This conclusion can be checked, observationally, by deriving a better estimate for the local value of $\partial^2 N/\partial z \partial W_0$ (especially at the high-$W_0$ end of the distribution which would be most affected by the extinction bias), and by carefully searching for background sources behind crowded stellar fields so that their density may be compared to that of objects in the field. A better understanding of the opacity of star-forming disks to background radiation would also reveal the degree to which our parameter estimates are affected by extinction bias at the high-$W_0$ and high-$\Upsilon$ tails of the distributions, and would help to determine whether additional physics (such as a luminosity dependent $f_L,\xi_w$) is required.

\subsection{Starburst Winds and \ion{Mg}{2} Systems' Kinematics}

Our model implies that the number counts of strong intervening systems can be fully accounted for by the number of  star forming disks [$f_L\sim 1$ given the number counts of \citet{maz98}] if the cross-section for strong \ion{Mg}{2}\ systems is of order the area of the \ion{H}{1}\ disk ($\xi_\sigma\sim1$). This allows us to calculate the mean \ion{Mg}{2}\ absorption line profile that would be observed against the galactic continuum for an ensemble of galaxies (e.g., with a given star formation rate) by stacking their individual (predicted) spectra.  Specifically, using the luminosity function of galaxies and their star-formation probability distributions, we created a mock catalog of galaxies and their associated absorbers according  to the modeling scheme of \S3. An absorption spectrum for each galaxy was then constructed under the assumption that the absorption trough is saturated and that the outflow fully covers the UV emitting disk [but see \citet{ma09}]. This means that we approximate the absorption line profile, as would be seen along a single sight line through a disk, as a rectangle centered at the systemic velocity of the galaxy. The mean line profile for a sample of galaxies, with some given properties, was then calculated by taking the mean flux at every photon energy of all spectra. This procedure is qualitatively similar to that carried out by \citet{we08} for DEEP2 data.  To allow comparison with the \citet{we08} results, we calculated $v_{75\%}$,  the velocity at which 75\% of the flux is transmitted through the blue wing of the (mean) line profile, as a function of the SFR of the galaxies . 

\begin{figure}
\plotone{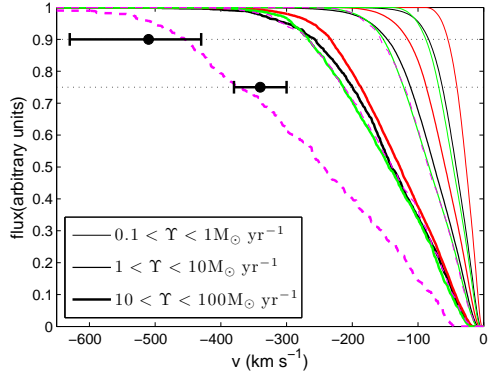}
\caption{Mean \ion{Mg}{2}\,$\lambda 2796$ absorption line profiles from galaxies with a specified range of star formation rates (see legend). Color coding of the different models follows that given in table \ref{tab}. In addition to models A, B, and C, we consider a model for which the UV emitting disk size corresponds to the core radius while the outer parts of the disk are UV faint and do not contribute to shaping the mean profile (magenta lines). The trough is symmetric but only the blue wing of the line is shown. The line shape is determined by an effective partial covering effect resulting from averaging over many systems, some of which having broader troughs than others (see text). Model B results in broader troughs on accounts of its higher $\xi_w$ value compared to model C. The results for the velocity at 75\% transmission (see horizontal dotted line) for model B, with its core radius being the size of the UV emitting disk, are in agreement with the measurements of \citet{we08} for galaxies with $10<\Upsilon<60\,{\rm M_\odot~yr^{-1}}$ whose mean $v_{75\%}$ and $v_{90\%}$ are shown as black points and full velocity range shown with error bars.}
\label{wei}
\end{figure}

The resulting line profiles are shown in Figure \ref{wei}. The shape of the predicted line profile, as would be observed along sight lines to background objects that pierce through intervening disks, is characterized by a deep trough at the systemic velocity of the galaxy with symmetrically extended blue and red wings. The shape of the mean line profile is not determined by optical depth effects --- individual systems are saturated by construction --- but rather by effective partial covering effects when averaging over the full sample of galaxies. This comes about from the fact that some systems in the ensemble have high velocity absorption components while others do not. In the latter case the unattenuated continuum contributes to the mean line profile. This predicts saturated mean absorption line troughs that do not reach a flux level of zero, consistent with the $1:1$ doublet ratio reported by \citet{we08} for their mean line profiles (note that our modeled absorption line profiles are not convolved with the instrumental line spread function hence the black troughs at the systemic velocity). 

The exact velocity up to which the line profile extends is not very sensitive to the chosen parameterization (solid line models in Fig. \ref{wei}). Nevertheless, for one of our runs, we assumed that the UV emitting region is of radius $\rho_0=\eta R_{\rm MgII}$, i.e., smaller than the size of the \ion{H}{1}\ disk. In this case only the outflow kinematics within $\rho_0$ will contribute to the formation of the mean line profile (magenta line in Fig. \ref{wei}). Models in which the outflow kinematics varies across the disk (e.g., model B) result in the absorption trough extending to much higher velocities. More quantitatively, for model B, we find that, for $10<\Upsilon<13\,{\rm M_\odot~yr^{-1}}$ ($30<\Upsilon<60{\rm M_\odot~yr^{-1}}$), $v_{75\%}\simeq 300\,{\rm km~s^{-1}}$ ($v_{75\%}\simeq 450\,{\rm km~s^{-1}}$), which is consistent with the results of \citet{we08} for \ion{Mg}{2}\ winds (see Fig. \ref{wei}). Results for models A, B, and C for which the UV emitting surface is the entire disk or the wind velocity is constant across the disk, seem to under-predict the observed characteristic outflow velocities. Our results imply that galaxies with a relatively low SFR, will show only low to moderate velocity outflows \citep{ru09}.

When observed at high spectral resolution, strong intervening systems often show several discrete absorption components, implying that the absorbing medium is clumpy. Drawing from the apparent correspondence between intervening \ion{Mg}{2}\ systems and galactic winds, this may indicate that galactic winds are largely clumpy and that their structure is characterized by clouds and filaments. Some support for this notion comes from \ion{Na}{1}\ studies of winds in star forming galaxies \citep{rup02,ma05,ma09,sat09}. Indeed, a highly complex density structure may be expected in such dense and dynamically evolving environments due to the development of various instabilities (e.g., Kelvin-Helmholtz) as the winds accelerate and expand \citep{st99,st04}. Conversely, the physical picture of galactic outflows naturally explains the rich line profile phenomenology of intervening systems: in this picture, the rich multi-component absorption that characterizes intervening systems is mostly due to (hydro-) dynamical effects related to outflow physics and is not related to the clustering of halos.  One way to test and refine our model would be compare the absorption line profiles of galactic winds, as observed toward the centers of star-forming disks, with those seen toward background objects whose sight lines pierce through the disks of the same galaxies. Even more revealing would be to find background objects behind star-forming galaxies and use their spectra to constrain the small scale physics of galactic winds, which is not accessible for studies using the galaxy itself as a background on accounts of the large emitting area \citep{ma09}.

\subsection{Redshift Evolution}

\begin{figure}
\plotone{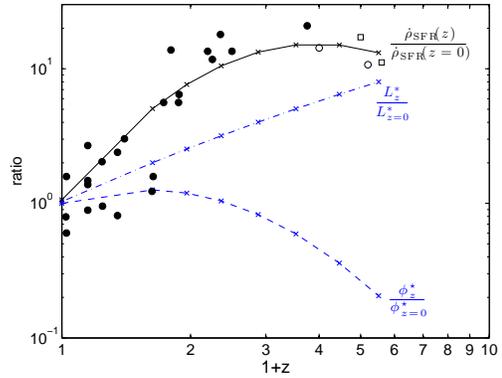}
\caption{Evolution in the statistical properties of galaxies with cosmic time. Shown in solid (dashed) line is $L^\star(z)/L^\star(z=0)$ ($\phi^\star (z)/\phi^\star(z=0)$) based on a second order polynomial fit to the $L^\star(z)$ ($\phi^\star(z)$) measurements of \citet{ga04} (for $z>0.45$) and \citet{ja08} (for $z\simeq 0$) in the $B$-band. If the luminosity of galaxies at any epoch is a measure of the star formation rate then the evolution in the star-formation rate density, $\dot{\rho}_{\rm SFR}(z)/\dot{\rho}_{\rm SFR}(z=0)$ is consistent with the observations. Data points for $\dot{\rho}_{\rm SFR}(z)/\dot{\rho}_{\rm SFR}(z=0)$ at $z<3$ were taken from \citet{ly07} and were normalized according to the measurements of \citet{ja08} at the local universe (filled circles). High redshift data were taken from \citet{hea04} based on data from \citet[empty circles]{ste96} and \citet[empty squares]{ou04}. Unsurprisingly, our parameterization for the luminosity function traces the global star formation rate density.} 
\label{evo0}
\end{figure}

\begin{figure*}
\plottwo{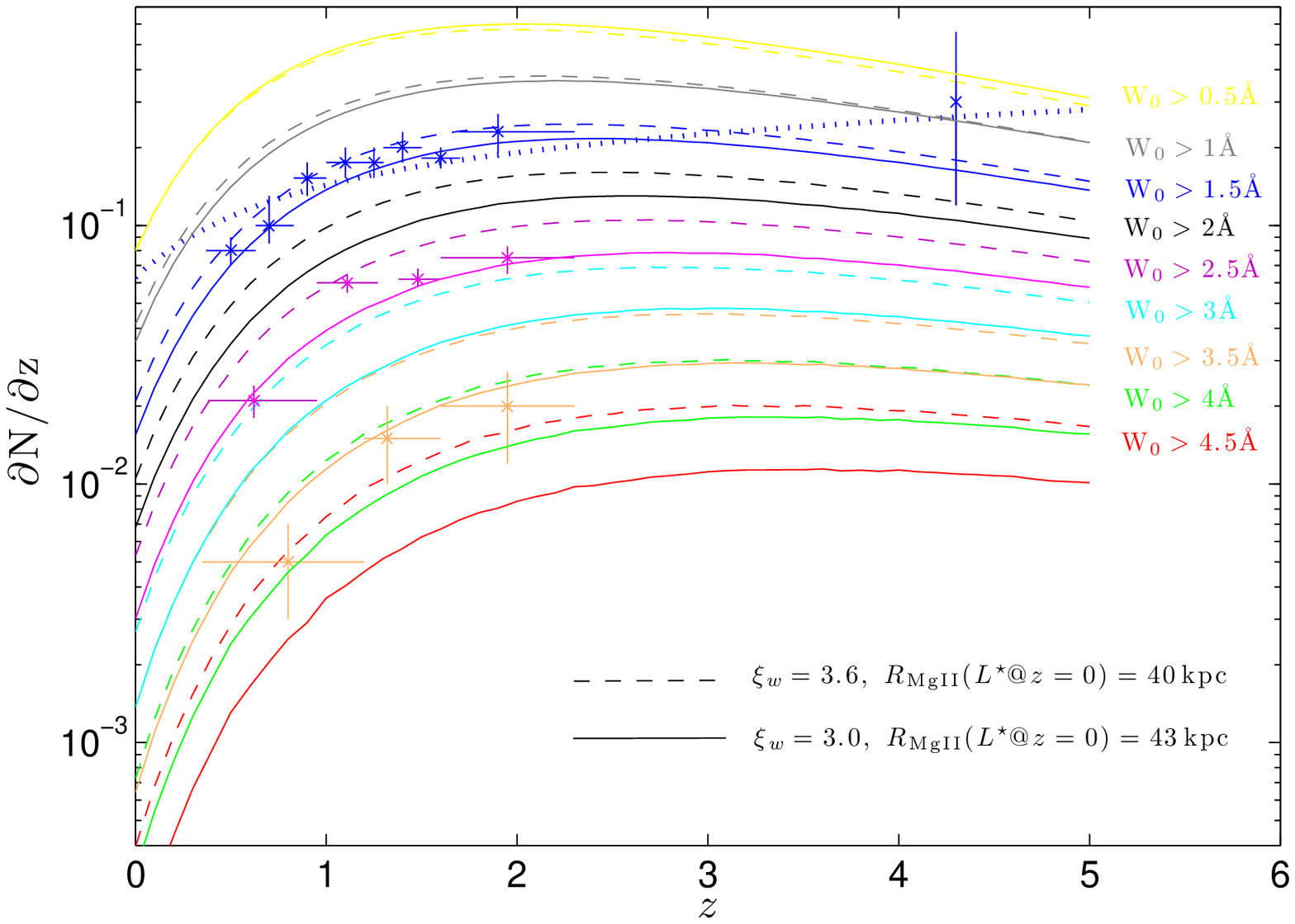}{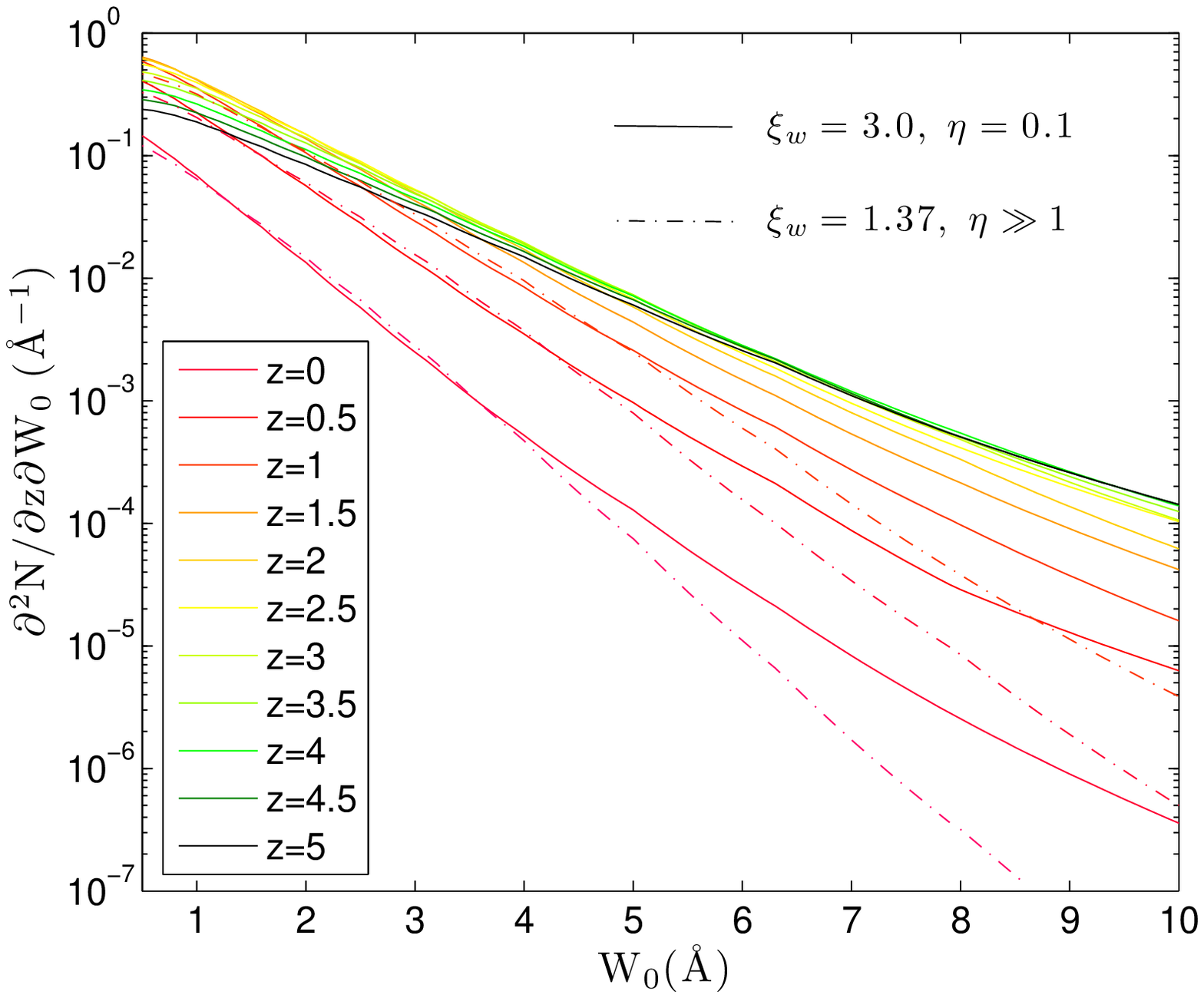}
\caption{Evolution in the statistical properties of strong intervening systems. {\it Left:} The $\partial N/\partial z$ evolution with redshift for several values of $W_0>W_0^{\rm lim}$ as denoted next to each curve. Also shown are data from \citet{nes05} for strong systems in the redshift range $0.4<z<2.3$. Two versions of model B are considered (solid and dashed lines) that are both consistent with the shape and amplitude of the signal for $W_0\gtrsim 1$\AA\ systems. In particular, all models predict a sharp rise in $W_0\gtrsim 1$\AA\ systems' density toward $z\sim 2$ followed by a gradual fall which follows from the decline in the global star-formation rate. Stronger systems number density peaks at higher redshifts. More specifically, our model B parameterization with $\xi_w=3.6,~R_{\rm MgII}=40$\,kpc (dashed line) over-predicts the signal for the strongest systems by 15-50\%. A better fit to current data over the full $W_0$ and redshift range is obtained for $\xi_w=3.0,~R_{\rm MgII}=43$\,kpc. The model predicts very different evolution track for strong systems than non-evolutionary models (dashed blue line for the case of $W_0>1.5$\AA\ systems. High redshift data are scarce and our model is consistent with current data [a point from \citet{jia07} for $z\sim 4.3$ systems is shown].  {\it Right:} Our modified model A (dashed line) and model B (solid line) predictions for $\partial^2 N(W_0,z)/\partial z \partial W_0$ for several redshift epochs (see color coding) shows evolution which is prominent for the strongest systems. In particular, the distributions functions flatten at high redshift showing the abundance of very strong systems at high redshifts. At low redshifts, the contribution of weaker systems (with $W_0\simeq 1$\AA) is relatively large while that of super-strong systems ($W_0 \sim 4$\AA) drops considerably, by about an order of magnitude, from $z=0.5$ to present times. Also shown is a departure from a pure powerlaw at the high-$W_0$ end of the distribution for our revised model B (as compared to our revised model A; dash dotted line), resulting from the contribution of sight lines passing close to the centers of relatively low luminosity galaxies. Testing which of the models is relevant requires better knowledge of the kinematics of galactic outflows across disks and better statistics for strong systems.}
\label{evo1}
\end{figure*}

Several authors have found indications for evolution in the number density of intervening systems as a function of redshift \citep{sar88,pet90,ste90}. Most relevant to our study are the recent results of \citet{nes05} who find tentative evidence for redshift evolution in the number density of absorbers, $\partial^2 N/\partial z \partial W_0$, as a function of $W_0$. In particular, evolution effects seem to be more pronounced for strong absorbers and the number of such systems decreases from high redshifts to present times [see, however, \citet{lun09} who find that the number density of $W_0>2$\AA\ systems actually increases from high redshift to $z\simeq 0.4$]. Nevertheless, the uncertainties in the observed evolution of $\partial^2 N/\partial z \partial W_0$ are large and better statistics, as a function of $W_0$, are required especially at low redshifts. In our model, $W_0$ is a direct consequence of the SFR of galaxies, which is known to be evolving. It is therefore natural to ask what our model predicts for the redshift evolution of $\partial^2 N/\partial z \partial W_0$ given the evolution in the properties of  galaxies over cosmic time.

In its present form, our model does not include evolutionary effects and additional assumptions are required to incorporate them into the formalism if any comparison is to be made with data relevant to absorption line systems at high-$z$. As noted in \S3, the properties of low luminosity star-forming galaxies are poorly determined at high redshifts. Nevertheless, using a specific set of assumptions we can check whether our model is consistent with observations. We rely on the recent work of \citet{ga04} who find that the Schechter luminosity function provides an adequate description of galaxy statistics at high-$z$, and give values for $\phi^\star$ and $L^\star$ for several redshift intervals. Due to the slope of the faint end of the high-$z$ luminosity function being difficult to determine due to incompleteness, we hold $\alpha$ fixed at the local value. (A similar approach was taken by Gabasch et al. only with $\alpha=-1.25$.) We fit $\phi^\star(z)$ and $L^\star(z)$ values reported by Gabasch et al. for $0.45<z<5$ together with the results of \citet{ja08} for $z=0$ by a second order polynomial so that:
\begin{equation}
\begin{array}{lll}
\displaystyle {\rm log} \phi^\star(z) & = & -2.65[{\rm log}(1+z)]^2+1.03{\rm log} (1+z)-2.52 \\
\displaystyle {\rm log} L^\star(z) & = & -0.30 [{\rm log} (1+z)]^2+1.42{\rm log} (1+z)+10.14
\end{array}
\label{lumiz}
\end{equation}
These relations, normalized to local values, are shown in Figure \ref{evo0}. 

Incorporating redshift evolution into our model is done in a straightforward way: we assume that all local relations and distributions hold at high redshifts and that only the luminosity function changes with cosmic time via equation \ref{lumiz}). Some support for our simplified approach may be gained from the findings of \citep{rav04} who found that the luminosity-dependent cross-section (or radius) of galaxies does not evolve with redshift in the range $0.25<z<1$, in agreement with the theoretical expectations of \citet{fal80}.  

If the characteristic SFR of an $L^\star$ galaxy is identified, on average, with its luminosity, at any epoch \citep{ja08}, then the predicted global star formation rate density of the universe, $\dot{\rho}_{\rm SFR}$, based on our parameterization for the luminosity function evolution with redshift, is consistent with current measurements \citep[and references therein; see Fig. \ref{evo0}]{ly07}. We caution, however, that some of the SFR measurements have recently been questioned and various systematic effects may exist in the estimation of $\dot{\rho}_{\rm SFR}(z)$ \citep{nor10}. The scenario defined above directly links the evolution in the statistical properties of strong intervening systems to that of star-forming galaxies and is therefore qualitatively different from the one advocated by \citet{ti09} in which absorbers' evolution is linked to the growth of dark matter halos over cosmic time.   

We integrated equation \ref{conv} to calculate $\partial N(W_0>W_0^{\rm lim};z)/\partial z$ for our model B for several values of $W_0^{\rm lim}$ --- see Figure \ref{evo1}. Our predictions for $\partial N/\partial z$ are very different from the expectations of non-evolutionary curves [$\propto (1+z)^2/E(z)$]. In particular, the decline in the number density of absorbers toward low redshifts is more pronounced than the one predicted by non-evolutionary curves by a factor of a few. This follows from the rapid decline in the star-formation rate from its peak value at $z\gtrsim 2$. At still higher redshifts, the mean star-formation rate density declines as does the predicted $\partial N(z)/\partial z$, which traces it, and is therefore very different than the predictions of the non-evolutionary curves (see Fig. \ref{evo1}). It is worth noting that a very different model, which ties the evolution of strong intervening systems to that of intermediate mass halos, was suggested by \citet[see their Fig. 6]{ti09} and predicts a much steeper decline in $\partial N/\partial z$ for $z>2$ than the one found here. Our calculations show that the density of absorbers above some $W_0$ limit peaks at intermediate redshifts, depending on $W_0^{\rm lim}$ (Fig. \ref{evo1}). Specifically, the density of $W_0\gtrsim 1$\AA\ systems peaks at $z\gtrsim 2$ while that of stronger systems with $W_0>3.5$\AA\ peaks at earlier times corresponding to $z\gtrsim 3.5$. At present, high-$z$ data are scarce and our model is consistent with recent measurements of the $\partial N(W_0>1.5{\rm \AA};z\simeq 4.3)/\partial z$ by \citet{jia07}. 

Comparing our model predictions to observations we see that the curvature in the data as a function of redshift is well traced by the models for a broad range of $W_0^{\rm lim}$. Furthermore, while the amplitude of the signal for $W_0>1.5$\AA\ systems is matched by the model, the model seems to over-predict the signal for $W_0>2.5$\AA\ by $\sim 15$\% and that for $W_0>3.5$\AA\ by $>50$\%. As noted by \citet{nes05}, the statistics of $W_0>3.5$\AA\ systems is poor and quantifying the systematics that affect the $\partial N/\partial z$ measurement is challenging. It is therefore possible that the true number density of such systems is actually higher than reported in their work. Whether this can account for the entire difference between the $\partial N/\partial z$ predicted by model B and the observations is unknown. Alternatively, our model, whose normalization relies on extrapolated data from high-$z$ measurements, may be inaccurate. As discussed in \S3, we have extrapolated the non-evolutionary curves of \citet{nes05} to $z=0$ to determine the relative normalization of the ${\partial^2N(W_0,z=0)}/{\partial z \partial W_0}$ curve while our model predicts a more rapid decline in the number of stronger systems at low-$z$. Attempting to fit the \citet{nes05} data set for $\partial N(W_0>W_0^{\rm lim};z)/\partial z$, we arrive to a somewhat different parameterization than our model B, which we term as model B1, with $\xi_w=3,~R_{\rm MgII}(L^\star(z=0))=43$\,kpc and with all other parameters held fixed at their model B values. This model shown in solid line traces well the shape and amplitude of the observed signal for a broad range of $W_0^{\rm lim}$. (We note that the model is also in good agreement with the other data sets considered in the previous sections.) Using model B1 to match the $\partial N(W_0>W_0^{\rm lim};z)/\partial z$ data of \citet{nes05} over a broad redshift range provides us with a refined estimate for ${\partial^2N(W_0,z=0)}/{\partial z \partial W_0}$ which is characterized by a lower density of the strongest systems that suffer from the most pronounced evolutionary effects at low redshifts; see figure \ref{dndw}. 

As already noted in \S3, predictions for ${\partial^2N(W_0,z=0)}/{\partial z \partial W_0}$ are expected to show deviations from a pure exponential for models of type B in which $\eta <1$. Studying the shape of ${\partial^2N(W_0,z)}/{\partial z \partial W_0}$ as a function of redshift shows that this is indeed the case and that the contribution of relatively faint galaxies to the high $W_0$-end of the distribution is most pronounced at $z=0$, occurring for $W_0\gtrsim 4$\AA\ systems where the  contribution of the relatively rare star-forming systems is small. At higher redshifts the deviation from a powerlaw is less pronounced yet may be possible to detect even at $z\sim 1$ provided the statistics of $W_0>5$\AA\ systems may be determined to better than $\sim 10$\%. For comparison we also plot a version of our model A with $\xi_w=1.37$ and all other parameters fixed at their model A values. This model provides a good fit to the data sets described above (aside from the \citep{we08} results; \S5.4) but does not show deviation from an exponential decay in ${\partial^2N(W_0,z)}/{\partial z \partial W_0}$ even at large $W_0$. Therefore, a test for the connection between outflows from star-forming disks and strong intervening systems would be to check for deviations from an exponential at high $W_0$ values at high redshifts and compare those to the mean \ion{Mg}{2}\  absorption troughs observed toward star-forming galaxies.  Nevertheless, any proper comparison between the data and the model requires a good understanding of potential systematics, such as dust extinction, that are expected to be more important at small impact parameters from galaxy centers, and which would select against detecting high $W_0$ systems.  

If this model is correct then by measuring the density of absorbers over a broad redshift range one may be able to measure  $\dot{\rho}_{\rm SFR}$ in a way which is complementary to that accomplished by studying the emission characteristics of star-forming regions. In fact, better measurement of $\partial N(W_0>W_0^{\rm lim};z\sim 4)/\partial z$ may provide interesting constraints on the SFR history of the universe and help constrain the physical model proposed here. It is, however, important to stress that, in order to be able to provide robust star formation density measurements, the physics of galaxies and their associated outflows needs to be better understood. Alternatively, should $\dot{\rho}_{\rm SFR}$ be estimated by other means, it may be possible to investigate the physics of bright as well as faint galaxies, in terms of their outflows and disk sizes, up to the highest possible redshifts (so long as a bright background object is present) thereby evading the limitations plaguing emission-based studies.

\subsection{Star Formation Rates for Strong Systems}

Several studies have aimed to measure the star formation rate associated with intervening absorption line systems \citep{wo03,ku06,wo06,bou07l,gh07,wi07,not09,nes10}. In some studies, only upper limits for the star formation rates of host galaxies were obtained, while in other, lower bounds were measured due to contamination by the quasar light, finite fiber or slit sizes, and dust extinction (assuming that the true host galaxy had been detected). In a few cases, integral field unit (IFU) observations were obtained leading to more reliable estimates for the SFR associated with strong systems. 

\begin{figure}
\plotone{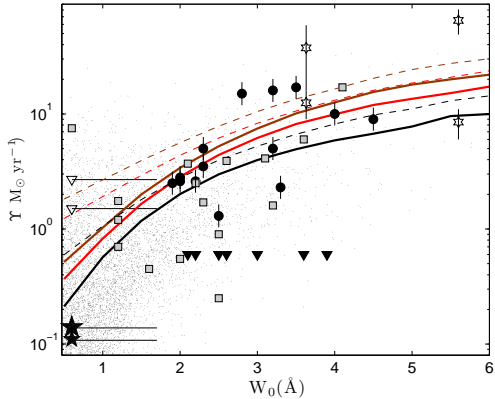}
\caption{Model predictions for the mean (solid line) and median (dashed line) star formation rate for $z=0$ (black curve), $z=1$ (red curve), and $z=2$ (dark red curve) models. For the $z=0$ model, a plot of the simulated systems is shown exemplifying the large scatter and the high SFR tail of the distribution resulting in the mean values being considerably higher than the median values, especially for low $W_0$. Also shown are measurements for individual absorbers from \citet{bou07l} whose average redshift is unity (filled circles and filled upper limit triangles). Empty triangles represent upper limit data for $z\gtrsim 2$ DLA systems from \citet{ku06}, for which $W_0$ is not known and a value of $W_0=0.6$\AA\ is assigned (see text). The uncertainty on $W_0$ in this case was calculated from the ${\partial^2 N(W_0,z=2)}/{\partial z \partial W_0}$ distribution and marks the $W_0$ range in which 68\% of the systems are expected to lie. Recent measurements from \citet{not09} are shown in gray squares. We also show data from \citet{wi07} which correspond to a statistical measurement at low ($0.4<z<0.8$; small star symbol) and high ($0.8<z<1.3$; large star symbol) redshifts, and for which we also assign a value of $W_0=0.6$ and calculate the uncertainty as before. New dust-corrected measurements for the star-formation in the environment of ultra-strong $z\sim 0.7$ systems by \citet{nes10} are marked as stars. Evidently, our model predictions are consistent with the data for the relevant redshift range. }
\label{bouche}
\end{figure}

We have used our model B1 at various redshifts, as defined above, to calculate the SFR distributions as a function of $W_0$. In particular, the results for a run of our model with $10^4$ absorber-galaxy pairs are shown in Figure \ref{bouche} where although most absorbing systems are associated with low star-forming galaxies, the scatter is large and a high star-formation tail is observed. This behavior is reflected in the large difference between the median and the mean of the SFR (as calculated from a run of our model B1 with $10^6$ absorber-galaxy systems) especially for $W_0\gtrsim 1$\AA\ absorbers. A clear trend is seen whereby the mean or median SFR increases with $W_0$. The mean and median values are also plotted for higher redshift models. These SFRs show a slight increase with $z$.  Below, we discuss some of the datasets used to derive SFRs at different redshifts, and examine how consistent they are with our models.

Measurements of the star formation rate for individual systems with known $W_0$ have been reported by \citet{bou07l}, using IFU observations, and are shown in Figure \ref{bouche}. These data correspond to $z\simeq 1$ systems with which our model predictions are consistent. Additional measurements have recently been reported by \citet{not09} for lower redshift ($z\sim 0.5$) systems and our model predicts comparable mean and scatter around the mean (compare the scatter in the simulated systems to the measurements in Figure \ref{bouche}). It should be noted that while the  \citet{not09} study is biased to galaxies with brighter emission lines, the reported SFR should still be treated as a lower limit on accounts of fiber losses.  Just how representative the \citet{not09} measurements are of the true SFR in absorber hosts is yet to be determined. Very high star formation rates around two sight lines with ultra-strong systems at $z\sim 0.7$ with $W_0>3$\AA\ were reported by \citet{nes10}, consistent with our model predictions. We note, however, that in these cases an interacting galaxy pair is observed and it is not clear which galaxy hosts the absorber and whether a particular galaxy-absorber association is at all meaningful in these cases.

\cite{ku06} placed upper limits on the star formation rate in $z\sim 2.4$ DLA systems of $\gtrsim 1\,{\rm M_\odot~yr^{-1}}$. As discussed in \S5.3, a large fraction of the DLA population may be associated with strong \ion{Mg}{2}\ systems.  Given the results of \citet{rao06} that show a rather uniform column density distribution for systems of different $W_0$ (for $W_0>0.6$\AA), the objects in the \citet{ku06} sample are likely to be associated with the more numerous $W_0\gtrsim 0.6$\AA\ systems [for $W_0<0.6$\AA\ systems the probability for detecting a DLA is considerably reduced given the data reported by \citet{rao06}]. Figure \ref{bouche} compares the upper limits of the \citet{ku06} sample, to which we have assigned a value of $W_0=0.6$\AA\ (and note that 68\% of selected systems are expected to lie in the range $0.6<W_0<1.7$\AA\ given the shape of the $\partial^2 N(W_0,z=2)/\partial z \partial W_0$ distribution; see Fig. \ref{bouche}), to the predictions for the star formation rate at $z=2$. Clearly, model predictions are in qualitative agreement with observations.

\citet{wi07} have measured the mean star formation rate for strong absorbers at intermediate ($z\sim 0.6$) and high ($z\sim 1.1$) redshifts (shown in Fig. \ref{bouche}). In particular, all systems with $W_0>0.6$ were considered in their study. As before, we assign a $W_0=0.6$\AA\ value and note that 68\% of those systems are expected to lie in the interval  $0.6<W_0<1.7$\AA. Our mean $z=0$ and $z=1$ models seem to over predict the data by almost an order of magnitude while the median values are within a factor $\sim 3$ of the measurements. A similar discrepancy was noted by \citet{wi07} when comparing their estimates for the SFR to those deduced by \citet{pr05} and \citet{rao06}. As pointed out by \citet{wi07}, fiber losses could be important, especially in cases where the host extends, or is located beyond, 8\,kpc from a quasar sightline. In fact, our model predicts that much of the absorption originates from scales as large as the size of the \ion{H}{1}\ disk, and hence beyond the central 8\,kpc of $>0.01L^\star$ galaxies. Therefore, the reported values should be considered as lower limits on the mean star-formation rates associated with strong systems. Lastly, we note a potential bias in probing highly star-forming systems using absorption line systems: the Kennicutt-Schmidt law predicts that, on average, disks with higher star formation rates would be more optically thick, having higher metallicities and, potentially, higher extinctions through their disks. It is therefore likely that the high SFR end of the distribution of absorbing galaxies will not be recorded in surveys where the background quasars are drawn from magnitude-limited samples. 

\subsection{The $W_0-\rho$ Relation for Strong Systems}

Our model gives specific predictions concerning the relation between $W_0$ and the impact parameter, $\rho$, from the center of the host galaxy (but not from satellite galaxies that may cluster around the true host). In comparing our model predictions to the \citet{ste92}, \citet{ste94}, and \citet{ste95} data set, we assume that these studies identify true absorber-host associations. The Steidel et al. sample includes the impact parameter between the absorber and the host galaxy, the rest equivalent width, and the luminosity of the host. To properly compare model predictions to the data, we first select a particular absorber-galaxy pair from the Steidel et al. sample. For the chosen system, which is characterized by a certain galaxy luminosity and impact parameter,  we calculated the predicted, $W_0=W_0(L,\rho)$,  according to our model B prescription, using a Monte Carlo approach. This procedure is then repeated for all absorber-galaxy pairs in the Steidel et al. sample\footnote{We do not show the results for models A and C but note their poor agreement with the data for the following reason: the Steidel et al. sample suggests a decline in the mean $W_0$ for $\rho>50$\,kpc. In contrast to that, models A and C are characterized by constant wind velocity across galaxy disks. As brighter galaxies have larger disks and more intense SFR, these models predict, on average, an increase in the mean $W_0$ with increasing impact parameter.}. 

\begin{figure}
\plotone{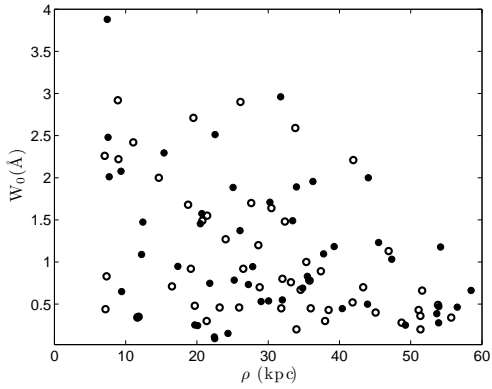}
\caption{One realization of our model predictions (see text) for the $W_0-\rho$ relation (black dots) and the data from Steidel et al. (circles). The match between model predictions and observations is satisfactory. It should be noted that the model does not include the contribution of satellite galaxies to the signal. More recent versions of the plot (not shown due to uncertainties on the hosts' luminosity) reveal strong absorption around galaxies on even larger scales than shown here.}
\label{chuck}
\end{figure}

The statistical predictions for $W_0(\rho)$ are compared to the observations in Figure \ref{chuck} in which a typical realization of the Monte Carlo simulation is shown. Clearly, the agreement is satisfactory in terms of the predicted mean and the scatter around the mean. Nevertheless, it is important to emphasize that one should be cautious to not over-interpret the outcome of this comparison.  In particular, to properly account for the observations, especially at large impact parameters, one needs to account for the possibility of satellite galaxies contributing to the absorption. This is currently not included in our model. While one does not expect the cross-section of satellite galaxies to fully cover the sky at large impact parameters around the host, we note that recent studies show that, in contrast to the classic picture, the covering factor for strong \ion{Mg}{2}\ absorption in galaxy halos is, on average, smaller than unity on $\lesssim 50$\,kpc scales \citep{tr05,ch08}. In particular, \citet{tr05} reported a covering factor of order 50\% for strong systems. If due to $\sim L^\star$ hosts then this value is qualitatively consistent with the expectations from randomly inclined thin disks.  A more comprehensive census of absorbers and satellite galaxies  on large scales around hosts is  needed to further test and constrain the model. 

\subsection{The Kennicutt-Schmidt Law Revisited}

In \S2 we qualitatively showed that strong intervening systems obey a version of the Kennicutt-Schmidt law for star formation. In particular, by combining observed relations between absorbers and star forming galaxies, the power-law dependence of the star formation rate on the (absorbing) \ion{H}{1}\ column density was deduced. In addition, the correct normalization of the Kennicutt-Schmidt law was obtained when using an order of magnitude estimate for the size of the \ion{H}{1}\ disk of absorbers' hosts. However, having now obtained a more quantitative connection between absorbers and galaxies, the proper disk size, as a function of $W_0$, may be used to arrive at a consistent Kennicutt-Schmidt relation for absorbers within the framework of the model. More specifically, we need to evaluate the expression
\begin{equation}
\Sigma_\star(\left < N_{\rm H} \right >)=\frac{\Upsilon[W_0( \left < N_{\rm H} \right >) ]}{\pi R_{\rm eff}\{ L[W_0({\left < N_{\rm H} \right > )] \}^2 }}.
\label{ss}
\end{equation}
Aside from $W_0(\left < N_{\rm H} \right >)$, all other relations are given by our model. In particular, $\Upsilon(W_0)$ and $L(W_0)$ are drawn from the distributions shown in Figure \ref{dist} whereas $R_{\rm eff}(L)$ comes from the fit to the $\partial^2 N (z=0)/\partial z \partial W_0$ distribution, as performed in \S4 for the different models considered here.

What $W_0(\left < N_{\rm H} \right >)$ relation should we take? The Kennicutt-Schmidt law, $\Sigma_\star(\left < N_{\rm H} \right >)$, connects the star-formation rate per unit surface in the disk to the {\it arithmetic} mean of the {\it total} hydrogen column density over the disk: $\left < N_{\rm H} \right >= \left < N_{\rm HI}+2N_{\rm H_2}\right >$.  For the \citep{ke89,ke98} samples, it was found that $\left < N_{\rm HI} \right > \simeq 2\left < N_{\rm H_2} \right >$ and we shall therefore assume that the total hydrogen column density, as traced by strong \ion{Mg}{2}\ systems, and as averaged over a finite patch in the disk, satisfies $\left < N_{\rm H} \right >= 2\left < N_{\rm HI} \right >$ \citep{fuk98}. Note that this does not necessarily imply that ${\rm H}_2$ molecules should be seen along sight lines to background quasars since galaxy disks are highly structured, the ${\rm H}_2$ gas has a low covering factor, and sight lines through molecular gas may be obscured by dust extinction \citep{zw06}. 

\ion{H}{1}\ observations of galaxy disks show a very broad range of column densities over the disk surface \citep{zw05}, and defining the relevant mean, $\left < N_{\rm HI} \right >$, is important. As discussed in \S5.3, the \ion{H}{1}\ distribution for galaxy disks is consistent with that measured for absorbers by \citet{rao06}. In particular, for both distributions, the arithmetic mean is much larger than the geometric mean. For strong intervening \ion{Mg}{2}\ systems \citet{me09} showed that the arithmetic mean is about $20$ larger than the geometric mean for $0.5<W_0<0.7$\AA\ systems in the \citep{rao06} sample. Interestingly, in that sample, the measured \ion{H}{1}\ column density for strong systems never exceeds $5\times 10^{21}\,{\rm cm^{-2}}$, even though the geometric mean over all systems in a given $W_0$ bin does increase as, roughly, $W_0^2$ (Eq. \ref{nhi}). At such a column density, the expected reddening for a galactic dust-to-gas ratio is $E(B-V) \sim 0.3$\,mag, and the SDSS is expected to miss most of the systems resulting in an under-estimation of $\left < N_{\rm HI} \right >$, perhaps explaining the flat $\left < N_{\rm HI} (W_0) \right >$ dependence \citep{me08}. If galaxy disks are self-similar, as would be expected for e.g., a turbulent medium down to the dissipation scale then the median and mean would, in fact, be proportional for all galaxy luminosities. For these reasons, we shall assume the following expression linking the {\it arithmetic} mean of the {\it total} hydrogen column density and $W_0$\footnote{We note, for completeness, that we have tried a different set of models in which the column density distribution along sight lines piercing through disks at certain annuli obeys   the \ion{H}{1}\ column density distribution of \citet[see their Fig. 3]{zw05} while the mean column density at any given annuli  decreases exponentially with the impact parameter from the galaxy center [e.g., \citet{kal08} and references therein]. This model resulted in a similar Kennicutt-Schmidt law as the one presented here and will not be further discussed due to its more complicated nature.}: 
\begin{equation}
W_0(\left < N_{\rm H} \right >)= \left ( \frac{\left < N_{\rm H} \right >}{1.2\times 10^{21}\,{\rm cm^{-2}}} \right )^{0.6}\,{\rm \AA}.
\label{wn}
\end{equation}
In doing so we implicitly assume that the \citet{rao06} data are of relevance to $z=0$ systems. This assumption is supported by our findings (\S5.3) and the findings of \citet{zw05} concerning the match between the \ion{H}{1}\ cross-section of $z=0$ galaxy disks and that required to account for high-$z$ DLA systems. 

\begin{figure}
\plotone{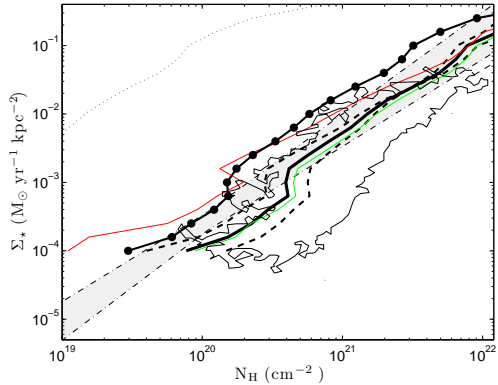}
\caption{Model predictions for the Kennicutt-Schmidt law as deduced from model B parameterization while assuming that the disk area over which star formation takes place is given by $\pi \rho_0^2$. In constructing the plot we use equations \ref{wn} \& \ref{ss} and plot model B as connected black points.  Clearly, the predicted Kennicutt-Schmidt relation for absorbers in the qualitative agreement with the observed Kennicutt-Schmidt law of galaxies [shaded area including uncertainties reported by \cite{ke98}]. Nevertheless, a small correction factor of about factor $\sim 2.5$ along the $x$-axis or a factor $\sim 0.3$ along the $y$-axis leads to a better quantitative match with the observations (solid black line). All models, A-C, give qualitatively consistent predictions (see table  \ref{tab} for color coding) upon multiplying their $N_{\rm H}$ values by a factor of $2.5$. Model B's predictions show also the uncertainty interval in dashed black curves. Recent measurements  by \citet{bi08} (thin contour lines) are also plotted showing a departure from a simple powerlaw for the Kennicutt-Schmidt law for low $N_{\rm H}$. Also shown in dotted line is our model predictions when the relation between $W_0$ and the geometric mean (instead of the arithmetic mean) \ion{H}{1}\ column density is used and assuming $\left < N_{\rm H} \right >=\left < N_{\rm HI} \right >$.}
\label{schmidt}
\end{figure}

The relevant radius, $R_{\rm eff}$, to consider in equation \ref{ss} is that which roughly corresponds to the size of the optical disk and over which the Schmidt-Kennicutt law was originally evaluated. In our model: $R_{\rm eff}(L) \simeq  \eta R_{\rm MgII}(L^\star) (L/L^\star)^{\gamma/2}\simeq 4(L/L^\star)^{0.35}$\,kpc, and therefore corresponds, by definition, to the core radius discussed in \S4.1. We ran Monte Carlo simulations obtaining, as before, a mock absorber-galaxy catalog listing $[\Sigma_{\star},\left < N_{{\rm H}} \right >]$ for each object, and calculated the mean relation between the two quantities shown in Figure \ref{schmidt}. 

Comparing our model predictions for the Kennicutt-Schmidt law derived for absorbers to the classic Kennicutt-Schmidt law for galaxies, we find that the slope of the relation matches well the galactic Schmidt-Kennicutt law. The obtained normalization has an offset of a factor $\simeq 1/3$ along the $x$-axis or factor $\simeq 3$ along the $y$-axis compared to the galactic Kennicutt-Schmidt law (see Fig. \ref{schmidt}). Such a discrepancy could result from several different reasons: it may be due to an under-estimation of the factor between the arithmetic and the geometric means of the \ion{H}{1}\ columns in strong systems, perhaps as a result of an extinction bias. Alternatively, the mismatch could be due to our adopted $R_{\rm eff}(L)$ being smaller by a factor $\simeq \sqrt{3}\simeq 1.7$ than the true radius. It may also be the result of our usage of high-$z$ relations between $W_0$ and $N_{\rm HI}$ to infer on the properties of local star-forming disks [a recent theoretical study by \citet{gk10} suggests that the Schmidt-Kennicutt law may be different at higher-$z$]. Nevertheless, given the uncertainties inherent in our analysis and the simplifications used, we find the match between the predicted Kennicutt-Schmidt law for absorbers and the observed one for galaxies to be very good.  Furthermore, upon taking, for example, the ratio of the geometric to arithmetic mean to be $50$, we find an excellent match with the Schmidt-Kennicutt law for galaxies.  A qualitatively similar agreement with the galactic Schmidt-Kennicutt law is obtained also for models A, B, and C (Fig. \ref{schmidt}).

To further check the correspondence between absorber and galaxies, it is of considerable interest to extend the $W_0$ range of the \citet{rao06} sample and quantify the Kennicutt-Schmidt law over a broader range of parameter space. Interestingly, deviations from the classic Kennicutt-Schmidt law at the low $N_{\rm HI}$ end have recently been reported for galaxy disks,  showing a flattening and a subsequent truncation at still lower column densities \citep{bi08}. Our model predictions suggest that most strong, $W_0 \gtrsim 0.5$\AA\ systems arise in faint galaxies having relatively low star formation rates. The nature of absorption systems is that they probe highly localized regions in space, so it is of interest to see whether deviations from a pure powerlaw are consistent between absorption and emission derived Schmidt-Kennicutt laws. If such deviations persist as better statistics are obtained then it may be possible to better quantify the correspondence between our view of disks in absorption and in emission. This will allow to probe faint star-forming disks at high redshifts using absorbers' phenomenology in a way which is complementary to that employed by emission studies. In particular, these approaches suffer from somewhat different systematics and their synergy will hopefully help to better quantify star formation phenomenon and galaxy evolution physics across cosmic time.

Thus far we have considered a global version of the Kennicutt-Schmidt law, as deduced from the phenomenology of \ion{Mg}{2}\ systems. Nevertheless, there may also be a local version of this law \citep{ke07} whereby local outflow properties relate to local star formation processes. At present we do not have observational evidence to support this hypothesis but note that a relation between the column density of \ion{O}{6}\ ions and the velocity dispersion of their resonance $1032$\AA\ absorption transition has been recently reported by  \citet[see also \citet{hec02}]{bo08}. Interestingly, the powerlaw slope of this relation is similar to the $N_{\rm HI}-W_0$ relation discussed in this work. It is possible that this is a different manifestation of the Kennicutt-Schmidt law, highlighting the preponderance of outflows and feedback effects from local star forming regions in the Galaxy.

\subsection{IGM Enrichment by Galactic Outflows}

We find no evidence that galactic outflows, as probed by strong \ion{Mg}{2}\ systems, extend considerably beyond the \ion{H}{1} disk  of their hosts (i.e., $\xi_\sigma \sim 1$). This is consistent with a scenario in which most outflows that give rise to strong \ion{Mg}{2}\ systems are launched from compact regions in the disk and do not necessarily escape the galaxy and join the IGM. That said, our results do {\it not} preclude the possibility that galactic outflows reach the large scale galactic halo since, on these scales, the outflow opacity drops as the gas expands. Mapping $W_0\gtrsim 0.5$\AA\ systems around nearby galaxies of known disk inclinations will shed light on the geometry and origin of the gas and help to determine its role in galaxy formation and IGM evolution models. In addition, better estimates for the spatial distribution of $N_{\rm HI} \gtrsim 10^{18}\,{\rm cm^{-2}}$ will determine just how far above the disk the cool component of galaxy outflows extends. In any case, the connection established here between disk outflows and strong intervening systems implies, by means of number counts, that the phenomenon of gas acceleration and ejection from compact star-forming regions occurs over a broad range of galaxy luminosities and is not limited to (bright) starburst galaxies \citep[and references therein]{hec03}. 

A {\it Chandra} imaging survey of edge-on disks  by \citet{st04} presented evidence for extended X-ray emission at right angles from disks for objects with significant star formation rates. These authors found that while a galaxy like the Milky Way shows extended X-ray emission on a few kpc scales above and below the disk plane, (two) fainter galaxies in their sample, with a lower star formation rate, showed no evidence for such emission. This led \citet{st04} to conclude that extended emission due to large-scale hot gas bubbles expanding from the disk only occurs when $\Upsilon>0.1\,{\rm M_\odot~yr^{-1}}$. Nevertheless, to explain the shape of the $\partial^2N/\partial z\partial W_0$ distribution for $W_0 \gtrsim 0.5$\AA, our model requires that outflows occur in disks with a star formation rate as low as $\Upsilon>10^{-2}\,{\rm M_\odot~yr^{-1}}$. Specifically, in our model, galaxies with $\Upsilon>0.1\,{\rm M_\odot~yr^{-1}}$ can account for the \citet{nes05} data for $W_0> 1$\AA. It remains to be seen whether the smooth shape of the $\partial^2N/\partial z\partial W_0$ distribution (with no apparent break at $W_0 \simeq 1$\AA) is due to a fortuitous  combination of strong- and weak-system populations. Alternatively, the absence of hot gas around galaxies with low SFRs could indicate that there is little physical connection between the cold gas that is traced by \ion{Mg}{2}\ systems seen to be outflowing from the disk, and the extended X-ray emitting gas extending high above the disk in some galaxies. If true, then this may indicate that cold gaseous outflows, manifested as strong \ion{Mg}{2}\ systems, are relatively compact, although covering most of the disk surface, and little can be inferred from their observable properties on the ultimate fate of galactic winds. It may also mean that while cool outflows are ubiquitous in star-forming systems, only in some objects are conditions favorable for launching large-scale X-ray emitting winds \citep{ev08}.  A better understanding of galactic winds or fountains may be gained by investigating the statistical properties of high-ionization line absorbers \citep{sem01,gr09}, and through the modeling of energy and momentum driven outflows \citep{st00,mu05,da07,fan07,ko07,ev08,fuj09}. 

It is worthwhile comparing the outflow velocity of cool \ion{Mg}{2}\ outflows to the circular velocity, $v_c$, of galaxies which serves as a measure of the escape velocity, $v_e \simeq \sqrt{2}v_c$. \citet{cou07} have quantified the dependence of the circular velocity , $v_c$, of galaxies on their luminosity and find that $v_c\propto L^{0.3}$, which is similar to the scaling of the wind velocity with the SFR. Noting that $L^\star$ galaxies at $z\sim 0$ produce stars at a rate of $\sim 1.7\,{\rm M_\odot~yr^{-1}}$, and using equation \ref{vw} we find
\begin{equation}
\frac{v_w}{v_e} \sim 0.5\xi_w,
\end{equation}
which implies that for our model A, $v_w/v_e \simeq 0.8 <1$ while for our model B, $v_w/v_e \simeq 1.8>1$ within $\rho_0$ (see \S5.4). We emphasize that these values for $v_w/v_e$ values are maximal in the sense that most galaxies will exhibit  slower \ion{Mg}{2}\ outflows according to the distribution defined in \S3. Therefore, if model A is relevant then \ion{Mg}{2}\ outflows probably do not escape galaxies and likely form galactic fountains instead. Model B is our more favored model since it accounts for the observed kinematics of galactic outflows (\S5.4), explains the dependence of $W_0$ on the impact parameter from the galaxy center, $\rho$, and is theoretically more plausible (\S4.1). In this case, judging from the outflow velocity distribution of Figure \ref{Ps}, we estimate that about  40\% of all galaxies or, alternatively, all galaxies for 40\% of the time, will eject outflows that could escape their potential wells (note that these estimates are sensitive to uncertain inclination corrections which we ignore in the present study).  

We can estimate the mass loss rate that may be associated with such outflows assuming they are able to escape the galaxy. Specifically, the ratio of the mass loss rate to the SFR for a galaxy of luminosity $L$ may be written as
\begin{equation}
\frac{\dot{M}(L)}{\Upsilon(L)} \sim \Upsilon^{-1} \pi \eta^2 R_{\rm MgII}^2 \frac{N_{\rm HI}}{r_h}m_p v_w\sim \xi_w r_{h,300\,{\rm pc}}^{-1}
\label{mdot}
\end{equation}
where we have taken a simplified disk geometry of radius $\rho_0=\eta R_{\rm Mg II}$ and assumed a constant velocity outflow being launched perpendicular to it. As the outflow velocity declines with $\rho >\rho_0$, resulting in a wind velocity smaller than the escape velocity, we neglect the contribution to the mass loss rate from radii larger than $\rho_0$. In estimating the mass loss rate, we have taken $r_{h,300\,{\rm pc}}=r_h/(300\,{\rm pc}) =1$ to be the relevant pressure scale height of the disk \citep{fer98}, and a typical hydrogen column for the outflowing material of $10^{19}\,{\rm cm^{-2}}$, as appropriate for $W_0\sim 1$\AA\ systems \citep{rao06} and assuming hydrogen is mostly in \ion{H}{1}. We also take a typical (one-sided) outflow speed of $v_w= 70\xi_w\,{\rm km~s^{-1}}$ (corresponding to a $W_0\simeq 1.4\xi_w $\AA\ system), with $m_p$ being the proton mass. We expect equation \ref{mdot} to have only a very weak dependence on galaxy luminosity since $R_{\rm MgII}^2v_w/\Upsilon \propto L^{0.7+0.3}/L\propto$\,const., which is consistent with the conclusions of \citet{dal07}. Assuming a 40\% duty-cycle for such outflows (see above) and relevant $\xi_w\gtrsim 1$ then galaxies expel gas from their centers at a rate comparable to the rate at which gas is turned into stars.  Interestingly, the typical density of the outflowing cool material used here is $N_{\rm HI}/r_h \sim 10^{-2}\,{\rm cm^{-3}}$, somewhat lower than the density of the cool ISM phase in the central regions of our galaxy but in good agreement with expectations on larger, $>10$\,kpc, scales \citep[see her Fig. 2]{fer98}. Clearly our estimates are very qualitative as the particular geometry and structure of the outflowing material is not known [c.f. \citet{tr07} who interpret similar \ion{Mg}{2}\ outflows in post-starburst galaxies as extending out to 100\,kpc from galaxy centers in the form of thin shells]. In addition, we do not know what fraction of the gas column density is bound to the disk and what remainder constitutes the outflow, which could lead to our above expression over-estimating the mass outflow rate by a factor of a few (assuming solar metaliclity) for a given pressure scale height of the disk. However, despite our qualitative estimates and using the phenomenology of strong intervening absorption systems, we arrive at a mass loss rate that is of order the star formation rate from galaxies, as has been suggested by other works based on very different sets of  arguments \citep{mu05,bou07m,dal07,da07}.

\section{Summary}

We have argued that quasar sight lines that show strong \ion{Mg}{2} absorption systems (with $W_0>0.5$\AA) in their spectra, largely probe intervening star forming disks. While a considerable fraction of the (\ion{H}{1}) column density associated with those  systems is likely to originate in the gravitationally bound star forming disk itself, the rest equivalent width of the lines, $W_0$, is largely set by the outflow kinematics.  A version of the Kennicutt-Schmidt law is obtained when combining the statistical properties of strong intervening systems with those of outflowing material from star forming galaxies. In this case, the observed cosmic density of intervening systems, $\partial^2 N/\partial z \partial W_0$, naturally follows from the luminosity function of galaxies and allows us to establish a statistical relation between absorbers and their host galaxies. Conversely, knowledge of the $\partial^2 N/\partial z \partial W_0$ distribution and the star-formation rate in galaxies may be used to shed light on the properties of galactic outflows with preliminary results pointing to an agreement between model predictions and kinematic data for \ion{Mg}{2}\ outflows.

The model presented herein gives detailed and testable predictions for the association between galaxies and absorbers. For example, based on a fit to the number density distribution of strong intervening systems, the model suggests that galaxies brighter than $>10^{-2}L^\star$ can give rise to strong intervening systems, which is in broad agreement with the observations. In addition we find that galaxies over a wide range of luminosities can give rise to absorbers of a given $W_0$. More specifically, $W_0\sim 1$\AA\ systems can originate from faint $L<0.1L^\star$ as well as bright $L>L^\star$ galaxies, which may help explain the lack of clear correlations between galaxy and absorber properties in samples with small numbers of quasar-galaxy pairs. Detailed predictions for the statistical relations between the $W_0$ of absorbers, and the star formation rate and luminosity of their hosts (and vice-versa) are provided and can be compared to observations. The model highlights a potential difficulty in the proper identification of the true host galaxies of absorbers due to their faintness and the presence of galaxy-galaxy clustering. This highlights the importance of UV observations of nearby intervening systems to properly study the absorber-galaxy connection and demonstrates the advantage in probing such the environments at high redshifts using GRBs. Further tests of our model are becoming possible with the recent installation of {COS} on board {HST}. In particular, it may be possible to obtain more reliable estimates for the local cosmic density of absorbers, $\partial^2 N(z=0)/\partial z \partial W_0$, which is crucial for appropriately calibrating the model and extrapolating its predictions to higher redshifts. Better identification of the true hosts of absorption systems and their associated star formation rate, particularly for the faintest galaxy hosts, will enable us to compare model predictions to observations. Mapping the outskirts of galaxy disks and looking for the signatures of galactic outflows may lead to a better understanding of galactic outflows with direct implications for intervening systems' phenomenology.

The simple correspondence advocated here between absorbers and star forming disks accounts for several additional properties of strong intervening systems. Specifically,  by using the luminosity-metallicity relation for galaxies and the deduced absorber-disk relation, we can account for the metallicity of absorbers extrapolated to $z=0$. Furthermore, considering the evolution in the luminosity function of galaxies with redshift, the evolution in the number density of absorbers per unit redshift, $\partial N(z)/\partial z$ is matched. Interestingly, in its simplest version, the model suggests that many of the statistical properties of star-forming galaxies (such as the specific star formation rate, the dependence of galaxy disk radius on luminosity, and the outflow velocity  dependence on the SFR) have not evolved, considerably, since $z\simeq 2$. As model predictions are rather sensitive to the scaling laws for galaxies, measurement of absorber density at higher redshifts may prove to be an important probe of galactic physics. In addition, we find that model predictions concerning the star formation rate in absorbers' hosts are consistent with current observations, yet uncertain systematic effects preclude a more comprehensive comparison at this stage. In addition, we find that the number density of Ly$\alpha$ absorbers with $N_{\rm HI}>10^{18}\,{\rm cm^{-2}}$ is broadly consistent with the results of 21\,cm emission studies of local galaxies. Nevertheless, the abundance of absorbers with  \ion{H}{1}\ columns in excess of $10^{21.3}\,{\rm cm^{-2}}$ seems to be exceed that which is inferred locally. At present we do not know whether this relates to evolution effects or to the very different beam sizes used by each approach. 

Attributing absorbers to gravitationally bound galaxy disks and their related outflows, avoids many of the theoretical complications associated with models that attempt to identify such systems with diffuse halo gas. For example, problems related to the formation of gas around galaxies that is cool, yet has a high velocity dispersion, are avoided. The need to advect metal-rich gas from the centers of galaxies to large scales without considerable mixing and disruption due to various  dynamical instabilities is also alleviated, and the relevant physics to consider is just that of the ISM. Cloud stability issues become irrelevant since the gas is, at least initially, gravitationally bound and a mass reservoir exists to replenish any evaporating fraction of it. Our findings point to a different physical state (and possibly also an origin) for weak \ion{Mg}{2}\ systems whose properties are not accounted for by the model. 

Our results imply that strong intervening systems can be used to study the properties of star-forming galaxies at high redshift where direct imaging, especially of faint objects, is challenging. In particular, our model implies that the evolution in the density of absorbers across cosmic time closely traces the global star formation rate density. Therefore, it may be possible to determine the star formation history of our universe using absorption line phenomenology. This nicely complements emission studies that are prone to different systematic effects. A further advantage is that absorption systems may be detected to very high redshifts provided enough bright background sources, such as quasars and GRBs may be found. In addition, metal absorption systems may be used to probe the outer parts of galactic disks, estimate their metallicity in an independent way, and assist in quantifying the metallicity gradient in disks. This has profound implications for galaxy formation and evolution, IGM enrichment history, and feedback processes across cosmic time. 

\acknowledgements

We thank M. Fumagalli, J. Hennawi, P. Khare, V. Kulkarni, B. M\'enard, and D. York for many enlightening discussions and the referee for thoughtful comments that considerably improved the paper. D. C. thanks Y. Gerber and A. Sternberg for continuous encouragement and support, and acknowledges many fruitful conversations with M. Balokovi\'c, M. Treselj, and D. Vinkovi\'c. 
We thank Chuli and FG for their continuing support. This research has been supported by a Long Term Space Astrophysics (LTSA) NASA grant NNG05GE26G awarded to D. V. B. and a Marie-Curie International Reintegration Grant PIRG-GA-2009-256434 awarded to D.C.

\end{document}